\documentclass[review]{elsarticle}

\usepackage{lineno,hyperref}
\modulolinenumbers[1]

\journal{Journal of NIM-A Templates}
\usepackage{graphicx}

\usepackage{bm}
\usepackage[table]{xcolor}
\usepackage{multirow}

\newcommand{\pp }{$p$+$p$ }

\begin{document}

\begin{frontmatter}

\title{Optical Transmission Characterization of Fused Silica Materials Irradiated at the CERN Large Hadron Collider}

\author[UI_phy,UI_npre]{S.~Yang}

\author[UI_phy,UI_npre]{A.~Tate}

\author[UI_phy]{R.~Longo\corref{mycorrespondingauthor}}

\author[CERN]{M.~Sabate~Gilarte} 

\author[CERN]{F.~Cerutti}

\author[CERN]{S.~Mazzoni}

\author[UI_phy]{M.~Grosse~Perdekamp}

\author[CERN]{E.~Bravin}

\author[BGU]{Z.~Citron}

\author[HQG]{B.~K\"{u}hn}

\author[HQG]{F.~N\"{u}rnberg}

\author[Columbia]{B.~Cole}

\author[UI_phy,UI_npre]{J.~Fritchie}

\author[BGU]{I.Gelber}

\author[UI_phy]{M.~Hoppesch} 

\author[CERN]{S.~Jackobsen}

\author[Maryland]{T.~Koeth}

\author[UI_phy]{C.~Lantz}


\author[UI_phy,FNAL]{D.~MacLean}

\author[Maryland]{A.~Mignerey}

\author[Kansas]{M.~Murray}

\author[CERN]{M.~Palm}


\author[UI_phy]{M.~Phipps}

\author[Kansas,ROM]{S.~Popescu}

\author[UI_phy]{N.~Santiago}

\author[BGU]{S.~Shenkar}

\author[Brookhaven]{P.~Steinberg}

\address[UI_phy]{Department of Physics, University of Illinois, 1110 W. Green St., Urbana IL 61801-3080, USA}

\address[UI_npre]{Department of Nuclear, Plasma and Radiological Engineering, University of Illinois, 1110 W. Green St., Urbana IL 61801-3080, USA}

\address[CERN]{CERN, CH-1211 Geneva 23, Switzerland}

\address[TUM]{Technical University of Munich (TUM), Department of Physics, Munich, Germany}

\address[BGU]{Ben-Gurion University of the Negev, Dept.\ of Physics, Beer-Sheva 84105, Israel}

\address[HQG]{Heraeus Quarzglas GmbH \& Co. KG, Kleinostheim, Germany}

\address[Columbia]{Columbia University, New  York,  New  York  10027  and  Nevis  Laboratories,  Irvington,  New  York  10533,  USA }

\address[Maryland]{University of Maryland, Dept.\ of Chemistry and Biochemistry, College Park, MD 20742, USA}

\address[Kansas]{University of Kansas, Dept.\ of Physics, Lawrence, KS 66045, USA}

\address[Brookhaven]{Brookhaven National Laboratory, Upton, NY 11973, USA}


\address[FNAL]{currently with Fermi National Accelerator Laboratory, Batavia, Illinois, USA}

\address[ROM]{On leave from IFIN-HH Bucharest, 077125, Romania}

\cortext[mycorrespondingauthor]{Corresponding author}

\begin{abstract}
The Target Absorbers for Neutrals (TANs) represent one of the most radioactive regions in the Large Hadron Collider (LHC). 
Seven $40$\,cm long fused silica rods with different dopant specifications, manufactured by Heraeus, were irradiated in one of the TANs located around the ATLAS experiment by the Beam RAte of Neutrals (BRAN) detector group. This campaign took place during the Run 2 \pp data taking, which occurred between 2016 and 2018. This paper reports a complete characterization of optical transmission per unit length of irradiated fused silica materials as a function of wavelength (240 nm - 1500 nm), dose (up to 18 MGy), and level of OH and H$_2$ dopants introduced in the manufacturing process. The dose delivered to the rods was estimated using Monte Carlo simulations performed by the CERN FLUKA team.
\end{abstract}


\end{frontmatter}


\section{Introduction}
\label{sec:introduction}


Fused silica materials are widely used in a variety of optical applications such as lenses~\cite{milam1998review} and telecommunications~\cite{miller2012optical} due to their excellent light transmission over a wide range of wavelengths, from the ultraviolet (UV) to the infrared~\cite{marshall1997induced, skuja2005defects, nurnberg2015bulk}. Fused silica is composed of pure silicon dioxide, SiO$_{2}$, in amorphous (non-crystalline) form
~\cite{brueckner1970properties, akchurin2003quartz}. The concentration of impurities like Al, Ca, Na, K, Mg, Ti is typically smaller than 0.015 ppm~\cite{silica_datasheet} in fused silica. Due to its high purity, fused silica has excellent radiation resistance against coloration compared to other glassy materials, such as fused quartz. The number of absorption sites is correlated with the concentration of impurities in the material, which trap charge carriers induced by radiation, resulting in the production of color centers~\cite{weeks1960trapped, griscom1991optical, griscom1996gamma}. For these reasons, fused silica is utilized in several applications designed to operate in radiation environments, including aerospace technology~\cite{salem2013transparent} and particle detectors at accelerator facilities~\cite{akchurin2003quartz,Bilki:8824611, akchurin2020cerium, abdullin2008design}. 

The CERN Large Hadron Collider (LHC) is the world's highest-energy particle accelerator~\cite{evans2008lhc}. With an upgrade of the injectors started during Long Shutdown (LS) 2, the accelerator officially began its transition towards the High-Luminosity (HL) era~\cite{hl_lhc, apollinari2017high}. The HL upgrade will be completed by the end of 2028, resulting in a higher collision rate, therefore increasing the radiation levels in the experiments and the accelerator tunnel. 
Some of the most critical regions of the accelerator, in terms of radiation, are the Target Absorber for Neutrals (TAN in the current LHC implementation, to be upgraded to TAXN for the HL-LHC \cite{PhysRevAccelBeams.25.053001}), which are the radiation absorbers for neutral particle debris produced by beam collisions in the ATLAS~\cite{aad2008atlas} and CMS~\cite{chatrchyan2008cms} interaction regions (IRs). Detectors that are installed and operated in the TA(X)N, including 
the Beam RAte of Neutrals (BRAN)~\cite{Matis:2016raz} and the Zero-Degree Calorimeters (ZDCs)~\cite{jenni2007zero, grachov2006status} of ATLAS and CMS, will experience unprecedented radiation levels in the HL-LHC. For this reason, a Joint Zero-degree Calorimeter Project (JZCaP)~\cite{atlas2021radiation} between the ATLAS and CMS ZDC groups was started to identify radiation-hard materials capable of withstanding the doses expected in the HL era. Given the similar challenges to be faced, the JZCaP and the BRAN group started a collaboration to study the radiation hardness of materials to be used for the HL upgrade of both 
detectors. Since both groups plan on constructing Cherenkov-based detectors, these studies were targeted at fused silica materials. 

The radiation characterizing the TA(X)N is unique since it is primarily due to high energy products generated in the showering of very forward neutral particles in the absorber. 
To study the radiation damage induced by this highly-energetic hadronic and electromagnetic cocktail, a new BRAN prototype detector, equipped with fused silica rods, hereafter referred to as  ``BRAN rods", was installed in one of the IR1 TANs during the 2016-2018 \pp run.

The BRAN rods were doped with different concentrations of hydroxyl (OH) and hydrogen (H$_2$) to study dopants' impact on the radiation hardness of fused silica. Given the nature of Cherenkov light, in particular its wavelength ($\lambda$) spectra, falling as $1/\lambda^2$, it is of great interest to study the radiation hardness of the radiator over a wide wavelength range extending into the UV region. 

The beam energy reached by the LHC in Run 2 (2016-2018) provided an opportunity to study the radiation hardness of fused silica over a wide dose range, up to $\sim$18 MGy. Detailed FLUKA~\cite{FLUKA:web, FLUKA:new, FLUKA:old} simulations were performed by the CERN FLUKA group, to estimate the dose accumulated in each part of the TAN, including the fused silica rods, during the Run 2 \pp irradiation campaign. FLUKA results enabled the possibility of correlating different rod segments with the doses accumulated in them.
The radiation damage was evaluated via measurements of the optical transmission of the irradiated fused silica samples compared to those of an un-irradiated sample (hereafter referred to as ``control"). In this manuscript, we report a multi-variate optical transmission analysis of irradiated fused silica, analyzing its dependence on wavelengths, doses, and material composition.

The paper is structured in the following way: Sec.~\ref{sec:BRAN} describes the BRAN prototype detector, the irradiation setup in the TAN and the corresponding FLUKA simulations used to estimate the dose accumulated in the BRAN rods. Sec.~\ref{sec:OpticalTransmittance} provides the details of the optical transmission measurements, while Sec.~\ref{sec:DataAnalysis} discusses the data analysis procedures. Results are presented and discussed in Sec.~\ref{sec:Results}. Finally, conclusions are given in Sec.~\ref{sec:Conclusions}.

%
%
%
%


\section{Irradiation setup of the BRAN prototype detector and FLUKA simulations}

\label{sec:BRAN}
\begin{figure*}[ht]
    \centering
    \includegraphics[width=1\linewidth,keepaspectratio]{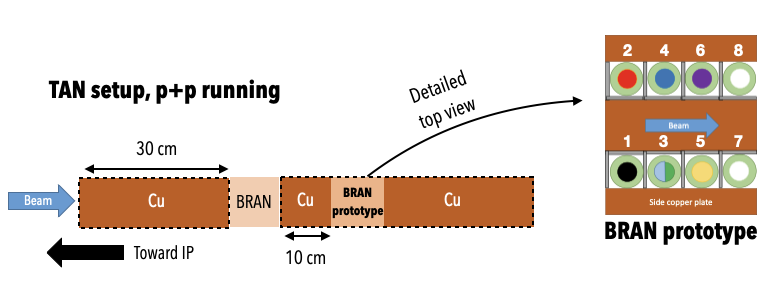}
    \caption{ 
    Schematic of the arm 8-1 TAN at the LHC during 2016-2018 \pp running. The numbers and colors in the right part of the figure identify the position of the rods in the BRAN prototype detector. Material specifications of the rods, as well as the maximum integrated doses accumulated in them, are shown in Tab.~\ref{tab:BRANrods}.}
    \label{fig:BRANSet}
\end{figure*}

Fig.~\ref{fig:BRANSet} shows a schematic of the BRAN prototype detector installed in the TAN on arm 8-1 of the ATLAS experiment during Run 2. The prototype detector was inserted in the TAN only for \pp running, and is constituted by three copper plates parallel to the beam propagation direction. A total of 8 slots are carved out in the copper, corresponding to a maximum of 8 fused silica rods to be accommodated. Each of the rods was manufactured by Heraeus Quarzglas ~\cite{heraeus_datasheet} and was 40 cm long, with a base diameter of 1 cm. 

The rods were characterized by different levels of OH and H$_2$ dopants, chosen to investigate the impact of material composition on the glass radiation hardness. Fused silica rods were inserted only into slot $1$ through $6$ (see Fig.~\ref{fig:BRANSet} for the layout), while positions $7$ and $8$ remained empty, allowing for the study of Cherenkov light yield in air (results not discussed in this paper). The specifications of each rod are listed in Tab.~\ref{tab:BRANrods}.

\begin{figure*}[t]
    \centering
    \includegraphics[width=1\linewidth,keepaspectratio]{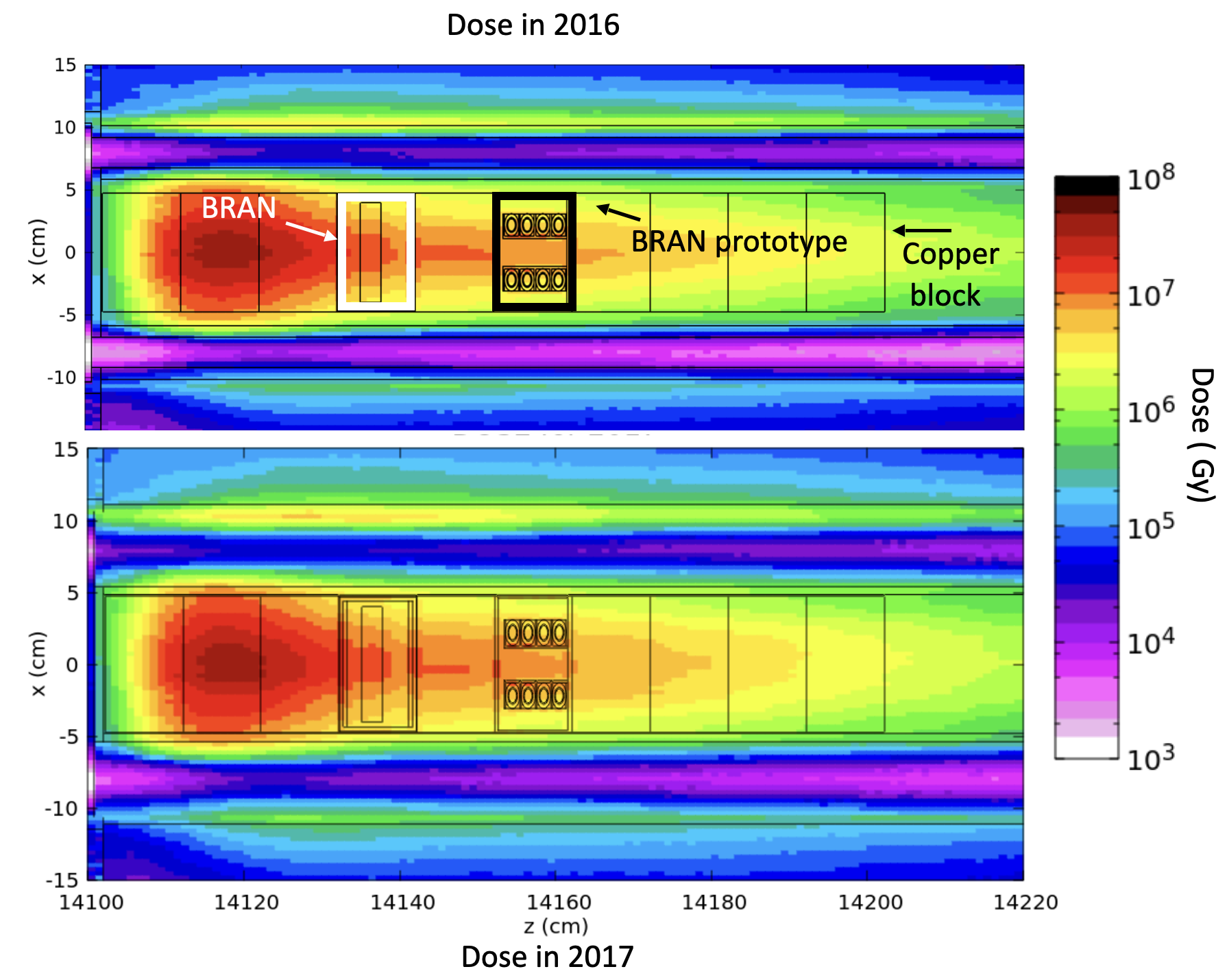}
    \caption{Accumulated dose $x$-$z$ profile in the TAN in 2016 (upper) and in 2017 (lower) \pp runs, respectively. The BRAN prototype is highlighted by the black box, while the white box marks the position of the BRAN detector. The integrated luminosity was 38.5 and 50 fb$^{-1}$ in 2016 and 2017, respectively.}
     \label{fig:Dose_profile_XZ}
\end{figure*}

\definecolor{mayablue}{rgb}{0.45, 0.76, 0.98}
\definecolor{lapislazuli}{rgb}{0.15, 0.38, 0.61}
\definecolor{meatbrown}{rgb}{0.9, 0.72, 0.23}
\definecolor{purpleheart}{rgb}{0.41, 0.21, 0.61}
\definecolor{pakistangreen}{rgb}{0.0, 0.4, 0.0}
\begin{table}[!htbp]
\footnotesize
\centering
\caption{Specifications of the irradiated fused silica rods. The number and color of each rod correspond to a given position in the BRAN prototype detector during the irradiation, as shown in Fig.~\ref{fig:BRANSet}. The same color scheme will be used when comparing results obtained from different materials. Rods 3a and 3b were placed in the same slot but in different periods.}
\vspace{0.2cm}
\begin{tabular}{|c|c|c|c|c|c|}
\hline
\textbf{BRAN}       & \textbf{Irradiation}  & \textbf{Max. Dose}    & \multirow{2}{*}{\textbf{Material}}    & \textbf{H$_{2}$}  & \textbf{OH} \\ 
\textbf{Position}   & \textbf{Period}       & [MGy]                    &                                       & [mol/cm$^{3}$]    & [ppm] \\ 
\hline
\hline 

\multirow{2}{*}{\textbf{Control}}    &  \multirow{2}{*}{None}    & \multirow{2}{*}{0}    & Spectrosil 2000   & \multirow{2}{*}{7.20e17}     & \multirow{2}{*}{1120} \\ 
                            &                           &                       & (High OH, Mid H$_{2}$) &                           &  \\ 
\hline

\cellcolor{black}        &  04/2016 -    & \multirow{2}{*}{18}    & Spectrosil 2000   & \multirow{2}{*}{7.20e17}      & \multirow{2}{*}{1120} \\ 
\multirow{-2}{*}{\cellcolor{black}\textbf{\textcolor{white}1}}   &   12/2018     &                       & (High OH, Mid H$_{2}$) &                               &  \\ 
\hline
\cellcolor{red}                                                 &  04/2016 -    & \multirow{2}{*}{10}      & Spectrosil 2000   & \multirow{2}{*}{7.20e17}      & \multirow{2}{*}{1120} \\ 
\multirow{-2}{*}{\cellcolor{red}\textbf{\textcolor{white}2}}   &   12/2017     &                           & (High OH, Mid H$_{2}$) &                               &  \\ 
\hline
\cellcolor{pakistangreen}                                                   &  04/2016 -    & \multirow{2}{*}{5}      & Spectrosil 2000   & \multirow{2}{*}{2.80e18}      & \multirow{2}{*}{1000} \\ 
\multirow{-2}{*}{\cellcolor{pakistangreen}\textbf{\textcolor{white}{3a}}}    &   12/2016     &                           & (High OH, High H$_{2}$) &                               &  \\ 
\hline
\cellcolor{mayablue}                                                &  04/2017 -    & \multirow{2}{*}{16}      & Spectrosil 2000   & \multirow{2}{*}{7.20e17}      & \multirow{2}{*}{1120} \\ 
\multirow{-2}{*}{\cellcolor{mayablue}\textbf{\textcolor{white}{3b}}} &   12/2018     &                           & (High OH, Mid H$_{2}$) &                               &  \\ 
\hline
\cellcolor{lapislazuli}                                                 &  04/2016 -    & \multirow{2}{*}{9}      & Spectrosil 2000   & \multirow{2}{*}{0}      & \multirow{2}{*}{1011} \\ 
\multirow{-2}{*}{\cellcolor{lapislazuli}\textbf{\textcolor{white}4}}   &   12/2017     &                           & (High OH, H$_{2}$ free) &                               &  \\                     
\hline
\cellcolor{meatbrown}                                               &  04/2016 -    & \multirow{2}{*}{8}      & Suprasil 3301   & \multirow{2}{*}{3.00e18}      & \multirow{2}{*}{15} \\ 
\multirow{-2}{*}{\cellcolor{meatbrown}\textbf{\textcolor{white}5}} &   12/2017     &                           & (Low OH, High H$_{2}$) &                               &  \\                     
\hline
\cellcolor{purpleheart}                                                 &  04/2016 -    & \multirow{2}{*}{17}        & Suprasil 3301   & \multirow{2}{*}{0}      & \multirow{2}{*}{14} \\ 
\multirow{-2}{*}{\cellcolor{purpleheart}\textbf{\textcolor{white}6}}   &   12/2018     &                           & (Low OH, H$_{2}$ free) &                               &  \\                     
\hline
\end{tabular}
\label{tab:BRANrods}
\end{table}

\begin{figure*}[ht]
    \centering
    \includegraphics[width=0.8\linewidth,keepaspectratio]{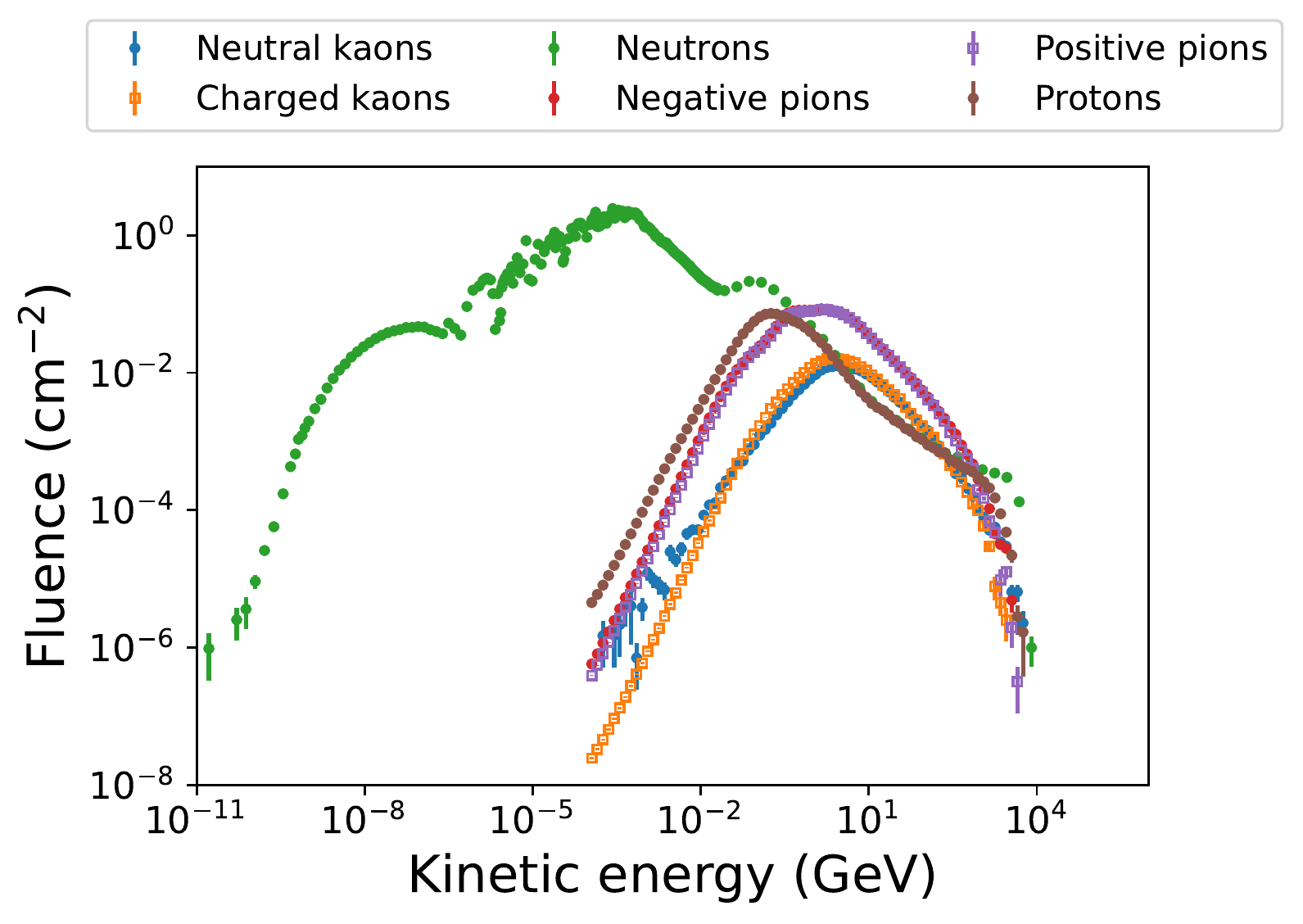}
    \caption{ 
Particle fluence spectra of different particle species impinging on the bottom-most 5 cm of Rod 1 in 2016, extracted from FLUKA simulations. In the 2016 crossing-angle configuration, this portion of the rod corresponds to the highest radiation levels registered in the material \cite{yang202222}. 
}
    \label{fig:particleFlux}
\end{figure*}

This work makes use of FLUKA simulations to evaluate the doses accumulated in the BRAN rods. 
FLUKA is a general-purpose Monte Carlo code for particle interaction and transport over a wide energy range. It has been benchmarked against recorded doses in the LHC and has been shown to have excellent agreement with data~\cite{alia2017lhc, Prelipcean:2777059}. Remarkable results were also achieved in the description of other aspects of the radiation environment in the accelerator, such as the activation of materials after Run 2 \cite{yang202222}. 
Using dedicated FLUKA simulations of the TAN region, it was possible to study the profile of the doses accumulated in the fused silica rods during Run 2. Because the detector experienced different beam crossing configurations in 2016 and 2017, two simulations were performed using the following settings:
\begin{enumerate}
    \item $p$+$p$ running in 2016: -180 $\mu$rad half crossing angle and integrated luminosity of 38.5 fb$^{-1}$. 
    \item $p$+$p$ running in 2017: +140 $\mu$rad half crossing angle and integrated luminosity of 50 fb$^{-1}$. 
\end{enumerate}
These values correspond to the integrated luminosity delivered to ATLAS during 2016 and 2017 \cite{ATLASluminosity}. 
Fig.~\ref{fig:Dose_profile_XZ} shows FLUKA results for the $x$-$z$ profile of the dose deposited in the TAN during \pp running in both 2016 and 2017. The right-handed coordinate system of the simulations is defined as follows: $x$ points outside the LHC ring in the horizontal plane, $z$ is along the ATLAS detector axis towards the right side, and $y$ is in the vertical direction towards the surface. The profile was extracted at the $y$ coordinate of the maximum dose value registered in the TAN. Note that, since 2017 and 2018 runs were characterized by the same beam crossing configuration, the dose map for 2018 was computed by rescaling the 2017 dose map using the ratio of the integrated luminosity delivered to ATLAS in the two years.
Thanks to FLUKA, it was also possible to characterize the spectrum of particles impinging on the rods during the irradiation time. An example is shown in Fig.~\ref{fig:particleFlux}. 

More details about the irradiation setup and the FLUKA simulations can be found in Ref.~\cite{yang202222}.  

%
%
\section{Optical transmittance measurements}
\label{sec:OpticalTransmittance}
The detailed dose profile of each rod provided by the FLUKA simulations enables the possibility of studying the transmittance of a given fused silica materials at different irradiation levels. As reported in Ref.~\cite{yang202222}, the dose accumulated in a single rod spans over four orders of magnitude along the vertical direction. Therefore, by cutting the rods into 1 cm segments, it was possible to form sub-sets of 40 samples, each characterized by the same material composition and different irradiation levels. 

A digital caliper was used to measure the maximum ($t^{max}$) and the minimum ($t^{min}$) length of a given sample. These lengths were determined by rotating the sample 360$^{\circ}$ within the caliper. The average cut length ($t_s$) was calculated as
\begin{equation}
 t_s = \left(\frac{t^{max}+t^{min}}{2}\right).
  \label{eq:avg_l}
\end{equation}
More details regarding the sample preparation can be found in Ref.~\cite{yang202222}.

\begin{figure}[!htbp]
\centering
\includegraphics[width=1\textwidth]{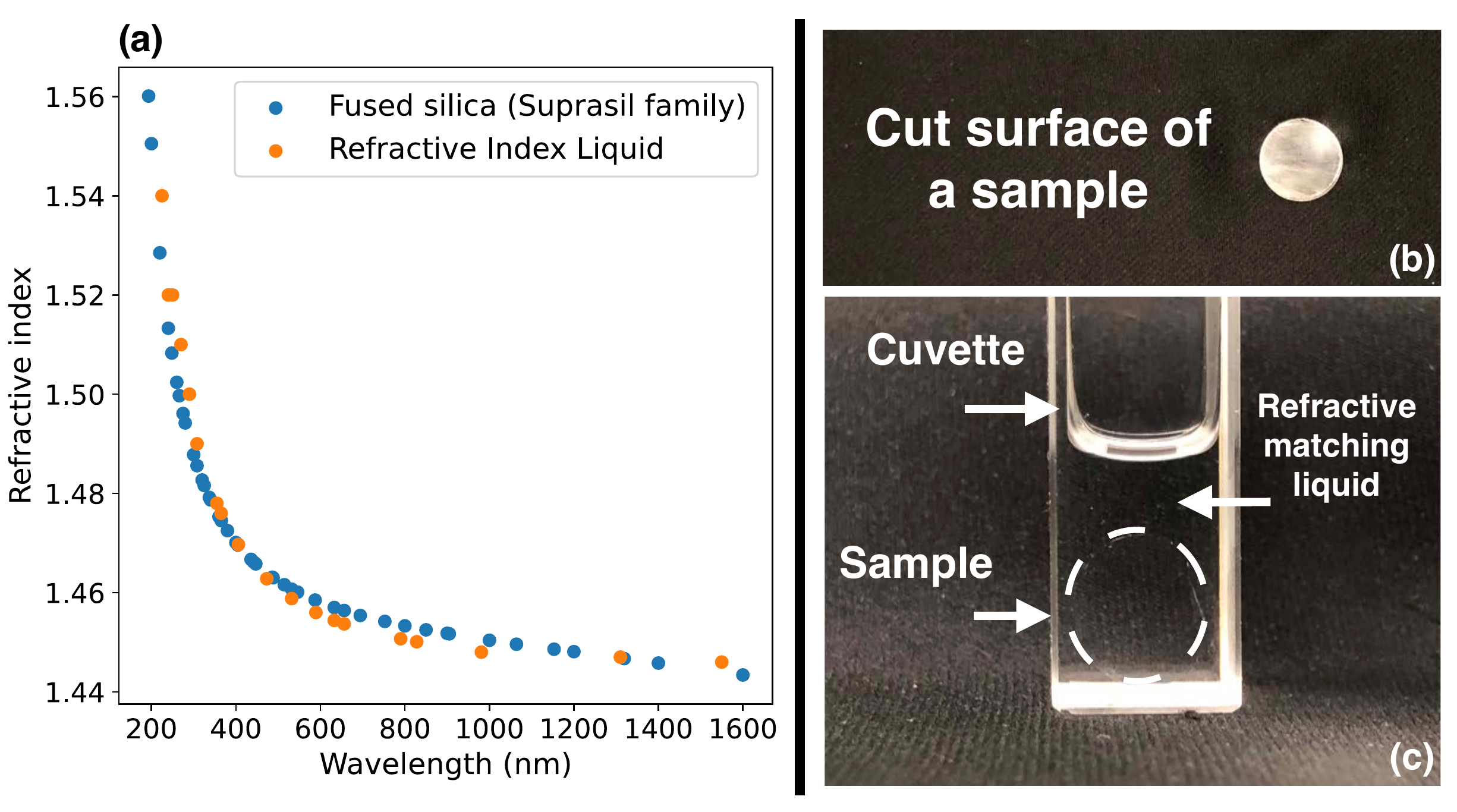}
\caption{(a) Refractive index comparison between fused silica (Suprasil family) and the chosen Series AA refractive index matching liquid. (b) The rough surface of a cut sample.  (c) The sample immersed within the refractive index matching liquid in the fused silica cuvette.}
\label{fig:refractive_index_comparison}
\end{figure}

During the cutting process, the blade introduces roughness on the cut surface.  
A refractive index matching liquid was used to reduce the effect of the surface on the the optical transmission measurements. Series AA liquid from Cargille Laboratories, characterized by a refractive index value of 1.456 at 589.3 nm at 25.0°C~\cite{index_liquid_datasheet}, was chosen for the measurement. A comparison between the refractive index of Series AA and fused silica is shown in the left panel (a) of Fig.~\ref{fig:refractive_index_comparison}. A Suprasil 3001 fused silica cuvette was used to contain the samples and the liquid during the measurements. The right panel of Fig.~\ref{fig:refractive_index_comparison} shows an example of a sample after the cut (b) and inserted in the cuvette with refractive index matching liquid (c).

\begin{figure}[!htbp]
\centering
\includegraphics[width=0.76\textwidth]{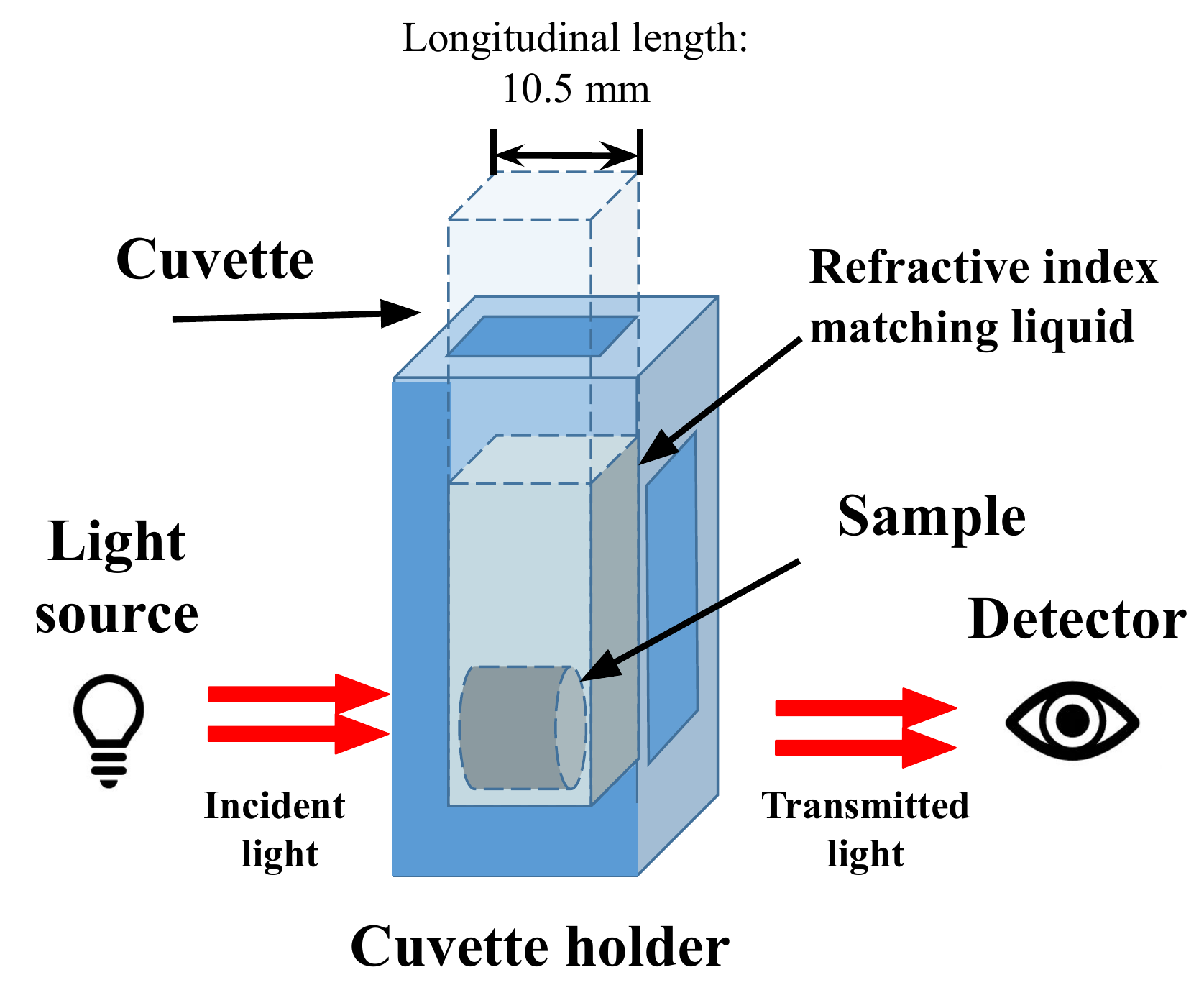}
\caption{Schematic drawing of the optical transmission measurement setup.}
\label{fig:Transmittance_setup}
\end{figure}

The transmittance of the samples was measured using a Varian Cary 5000~\cite{Cary5000} spectrophotometer, capable of performing measurements over a large wavelength range. This work presents results from 240 to 1500 nm. The sampling interval was 1 nm, the average measurement time was 0.1 seconds, and the spectral bandwidth was 2 nm. Air was used as a reference to calibrate the instrument for all the measurements presented in this study, including the fused silica samples and the refractive index matching liquid. For each measurement, the sample was immersed in the refractive index matching liquid in a cuvette placed in the Cary's built-in cuvette holder. The sample was centered with respect to the optical beam. A schematic of the experimental setup is shown in Fig.~\ref{fig:Transmittance_setup}.

%
%
\section{Data analysis}
\label{sec:DataAnalysis}
\subsection{Correction for the refractive index matching liquid attenuation}
\label{ssec:Attenuation_correction}

The Beer-Lambert law ~\cite{swinehart1962beer}, which describes the attenuation of light in a slab of absorber material with parallel faces, was used to correct for the effects introduced by the refractive index matching liquid. 
As the cuvette is filled with the refractive index matching liquid, the attenuation coefficient of the latter, $\alpha_l$, can be calculated as
\begin{equation}
    \alpha_l  =\frac{-1}{t_{c}} \ln(I_l),
  \label{eq:liquid_attenuation_coefficient}
\end{equation}
where $t_c$ is the longitudinal length of the cuvette and $I_{l}$ is intensity of the light measured\footnote{In this work, the average of ten different measurements was used to determine $\alpha_l$.} with only liquid in the cuvette. In this case, because no sample is placed in the cuvette, the longitudinal length of the liquid is equal to that of the cuvette. The longitudinal length $t_{c}$ of the cuvette used in this study is 1.05 cm.

To carry out one measurement, the given sample needs to be inserted into the cuvette. Therefore, the longitudinal length of the liquid $t_l$ reduces to 
\begin{equation}
   t_l = t_c - t_s,
  \label{eq:liquid_thickness}
\end{equation}
where $t_s$ is the length of the sample computed using Eq.~\ref{eq:avg_l}.
The corrected intensity, $\hat{I}$, is then calculated as
\begin{equation}
  \hat{I} = I e^{\alpha_{l}  t_l},
  \label{eq:attenuation_correction}
\end{equation}
where $I$ is the measured intensity, and $t_s$ and $\alpha_l$ correspond to the index matching liquid.
The transmittance $T$ of the sample was calculated by comparing the measured intensity to that of the control sample~\cite{hope2014diffuse}
\begin{equation}
    \label{eq:transmittance}
    T = \frac{\hat{I}}{\hat{I}_{u}},
\end{equation}
where $\hat{I}_{u}$ is the corrected intensity of the control sample, which was obtained by applying the aforementioned procedure (Eqs.~\ref{eq:liquid_thickness} - \ref{eq:attenuation_correction}) to measurements of the control sample. 

To allow for an unbiased comparison between different samples, the transmittance of each sample should be normalized to transmittance per unit length. 
To achieve this goal, one can compute the attenuation coefficient $\alpha_{s}$ of the sample as follows
\begin{equation}
    \alpha_{s}  =\frac{-1}{t_{s}} \ln(T).
  \label{eq:sample_attenuation_coefficient}
\end{equation}
and then use $\alpha_{s}$ to evaluate the transmittance per unit length $\bar{T}$ as:
\begin{equation}
    \bar{T} =  TC,
  \label{eq:length_correction}
\end{equation}
where C is a correction factor corresponding to
\begin{equation}
    C = e^{-\alpha_s (t_n-t_s)},
\end{equation}
In the last formula, $t_n$ represents the arbitrary unit length chosen for the normalization. Because the nominal length of the samples used in the transmission measurements is 1 cm, $t_n$ was set to 1 cm in this study. 

\subsection{Uncertainty estimation}
\label{ssec:Error_estimation}

This study identified two primary sources of uncertainty, one introduced by the attenuation correction described in Sec.~\ref{ssec:Attenuation_correction}, and the other related to the systematic error due to positioning and alignment variation of samples in each measurement. 

The uncertainty on the attenuation coefficient of the liquid $\sigma_{\alpha_{l}}$, defined in Eq.~\ref{eq:liquid_attenuation_coefficient}, was calculated as
\begin{equation}
    \sigma_{\alpha_l} = \sqrt{\sigma_{I_l}^2 \Bigl( \frac{-1}{I_l t_{c}} \Bigr)^2 + \sigma_{t_{c}}^2 \Bigl( \frac{\ln I_l }{t_{c}^2} \Bigr)^2},
    \label{eq:decay_coef_err}
\end{equation}
where $\sigma_{I_{l}}$ is estimated using the standard deviation of 10 measurements with only liquid in the cuvette, and $\sigma_{t_{c}}$ is the uncertainty on the length of the cuvette specified by the manufacturer, 0.05 mm.

The uncertainty on the corrected intensity calculated in Eq.~\ref{eq:attenuation_correction}, hereafter referred to as $\sigma_{\hat{I}}$, can be derived using standard error propagation:
\begin{equation}
    \sigma_{\hat{I}} = \hat{I}  \sqrt{ (\alpha_l  t_{l})^2 \Bigl [ \Bigl ( \frac{\sigma_{\alpha_l}}{\alpha_l} \Bigr)^2  + \Bigl (\frac{\sigma_{ t_{l}}}{ t_{l}} \Bigr)^2 \Bigr] + \Bigl( \frac{\sigma_{I}}{I} \Bigr)^2 },
\label{eq:final_corrected_I_err}
\end{equation}
where $\sigma_{I}$ is the uncertainty on the sample's intensity measurement and $\sigma_{t_l}$ is the uncertainty on the longitudinal length of the liquid in the measurement. The attenuation caused by the liquid is affected by the uncertainty on the liquid thickness, $\sigma_{t_l}$, that was assumed to be equal to
\begin{equation}
    \sigma_{t_l} = t^{max} -t^{min}.
\label{eq:liquid_error}
\end{equation}
This assumption provides a conservative uncertainty on the corrected intensity. Note that the Cary 5000 spectrophotometer does not provide the uncertainty on measurements so $\sigma_{I}$ was not available and assumed to be negligible. 
However, the largest contribution to $\sigma_{I}$ comes from variations between each measurement, including the sample positioning and fluctuations in the spectrophotometer performance.
In this analysis, such effects are accounted for by evaluating a systematic uncertainty associated with the transmittance results' reproducibility, which will be described later in this section.

The uncertainty on the transmittance of the sample, $\sigma_{T}$, can be expressed as
\begin{equation}
    \sigma_{T} = T \sqrt{\Biggl(\frac{\sigma_{\hat{I}}}{\hat{I}} \Biggr)^2 + \Biggl(\frac{\sigma_{\hat{I_u}}}{\hat{I_u}} \Biggr)^2}, 
\label{eq:transmittance_error}
\end{equation}
where $\sigma_{\hat{I}}$ and $\sigma_{\hat{I_u}}$ are the uncertainty on the corrected intensity of the sample and the control sample computed by using Eq.~\ref{eq:final_corrected_I_err}, respectively.
The uncertainty $\sigma_{\alpha_s}$ on the sample's attenuation coefficient was computed as  
\begin{equation}
    \sigma_{\alpha_s} = \sqrt{\sigma_{T}^2 \Bigl( \frac{-1}{T t_{s}} \Bigr)^2 + \sigma_{t_{s}}^2 \Bigl( \frac{\ln T }{{t_{s}}^2} \Bigr)^2},
    \label{eq:sample_decay_coef_err}
\end{equation}
where $\sigma_{t_{s}}$ is the uncertainty on the sample's length, assumed to be the same as $\sigma_{t_{l}}$. 
Finally, the uncertainty $\sigma_{\bar{T}}$ on the transmittance per unit length calculated using Eq.~\ref{eq:length_correction} was evaluated as
\begin{equation}
 \sigma_{\bar{T} } =  \sqrt{\sigma_{T}^2 C^2 + \sigma_{t_{s}}^2 (T C \alpha_{s} )^2 + \sigma_{\alpha_{s}}^2 (T C (t_{s}-t_{n}))^2}
 \label{eq:T_correction_coef}
\end{equation}

The systematic error resulting from variations introduced by sample positioning in each measurement due, for instance, to different rotations of the samples in the cuvette or thickness of liquid upstream/downstream of the sample, was also evaluated. A set of representative samples were selected based on their length and position in the rod, using the following procedure. First, the 40 samples obtained from each rod were grouped in batches of 10 consecutive samples each. Then, the samples characterized by the maximum and minimum lengths were picked within each of these groups, resulting in 8 samples for each 40 cm long rod. The control and each selected sample were measured ten times.
For each measurement, the sample's orientation within the holder was varied by randomly rotating and flipping the sample.

Every measurement of the selected sample was paired with ten control measurements to calculate the corresponding transmittance ($T$), generating 100 transmittance results per selected sample. At each wavelength, the maximum and minimum transmittance values among the 100 results were obtained for each selected sample, and then the difference between the two extremes was computed. For each rod, at every wavelength, the maximum difference among the eight samples was then taken as a conservative estimate of the systematic error and propagated together along with the attenuation error to compute the final uncertainty on the transmittance per unit length ($\bar{T}$). 



%
%
%
%
\section{Results and discussion}
\label{sec:Results}

This section first presents the measured transmittance as a function of dose and wavelength. A correlation between these two variables and the radiation damage experienced by the fused silica material, namely its transmittance degradation, is observed.
Then, subsets of the BRAN rods sample results are selected based on their H$_2$ and OH levels to analyze the dopants' impact on fused silica's radiation hardness at different dose levels. 
Note that the ``transmittance" mentioned in the following sections is synonymous with the transmittance per unit length ($\bar{T}$). 

\subsection{Transmittance degradation as a function of dose and wavelength}

Thanks to FLUKA simulations, it was possible to correlate the transmittance with the dose received by each rod segment. In the simulations, the whole geometry was divided using a mesh (reflected by the resolution in Fig.~\ref{fig:Dose_profile_XZ}). In each mesh element, the deposited energy and the corresponding error were computed using FLUKA. To calculate the dose of each rod segment, the dose registered in all the cells within a segment was averaged. The uncertainty associated with the average was computed using standard error propagation.

\begin{figure}[!htbp]
\centering
\includegraphics[width=1\textwidth]{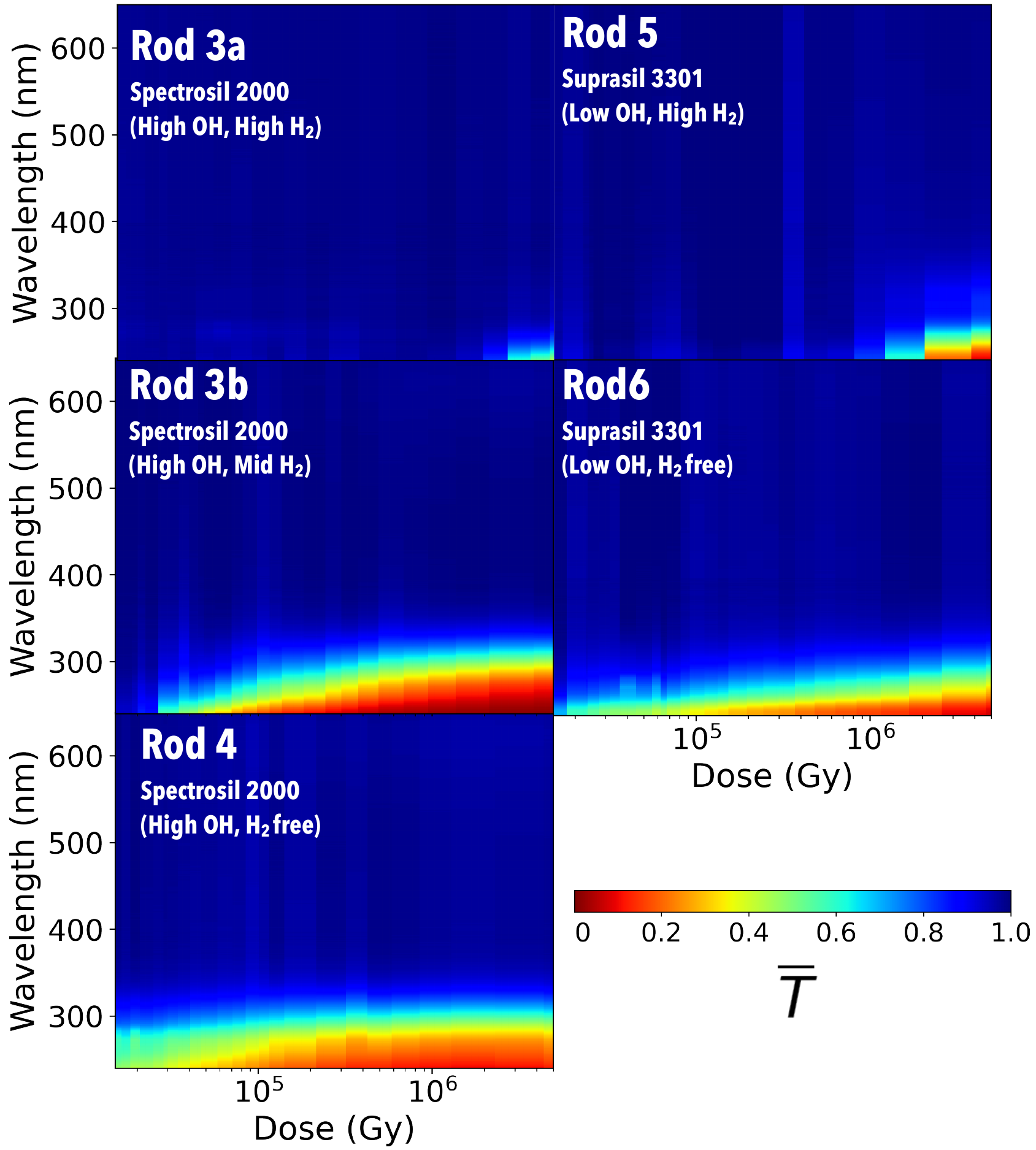}
\caption{Transmittance as a function of dose and wavelength for Rods 3a-6. The vertical axis represents the wavelength of the transmittance measurement. The horizontal axis displays the dose received by each sample, estimated using FLUKA. Note that the horizontal axis was limited to the dose range experienced by all rods, 1.5 $\times$  10$^4$ to 5 $\times$ 10$^6$ Gy, and the vertical was restricted to 240 - 650 nm. The upper limit was chosen in function of the typical primary sensitivity range of common photomultiplier tube's photocathode~\cite{hamamatsu2017photomultiplier}. } %
\label{fig:2D_spectral_T_vs_dose}
\end{figure}

\begin{figure}[!htbp]
\centering
\includegraphics[width=1\textwidth]{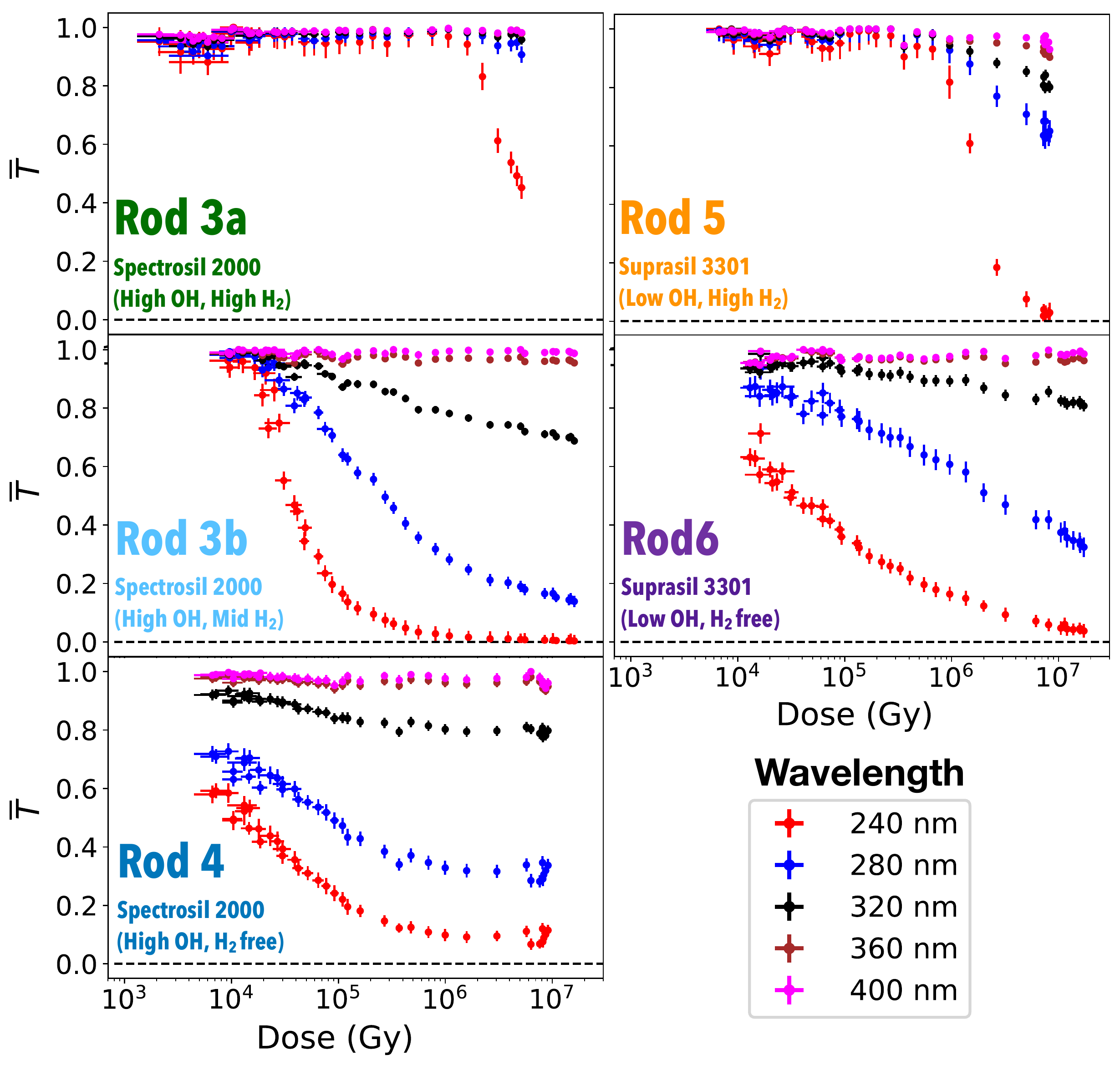}
\caption{Transmittance of Rods 3a-6 as a function of dose for five different wavelengths, in increments of 40 nm from 240 to 400 nm. The vertical axis represents the transmittance. The horizontal axis displays the dose received by each sample, estimated using FLUKA.}
\label{fig:UV_transmittance}
\end{figure}

Fig.~\ref{fig:2D_spectral_T_vs_dose} shows the transmittance for Rods 3a, 3b, 4, 5, and 6 as a function of the received dose and the wavelength. Among the different Spectrosil (high OH) rods available with an intermediate level of H$_2$ dopants, Rod 3b was chosen for the analysis because its position is consistent with that of Rod 3a. As expected, the transmittance decreases as the accumulated dose in the samples increases. In addition to the dose dependence, most transmittance degradation is observed in the UV region ($<$ 400 nm). Therefore, although the wavelength of the measurements goes to 1500 nm, the wavelength range in the figure was limited to a range of interest for typical photomultiplier applications, e.g. up to 650 nm. 
Fig.~\ref{fig:UV_transmittance} presents the transmittance versus dose in the UV region, where the most of the transmittance degradation is observed. 

The transmittance of each rod was plotted as a function of dose for five different wavelengths, in increments of 40 nm from 240 to 400 nm. Results at 360 and 400 nm show minimal attenuation, even at 10 MGy, suggesting an excellent radiation hardness of fused silica above these values independent of the dopant levels. The highest level of degradation is observed at 240 nm, the lower limit of the wavelength spectra, for all materials studied. However, distinct attenuation patterns can be observed for rods characterized by different dopant levels, suggesting an effect of OH and H$_2$ concentration on the fused silica's radiation hardness. Among all the rods, Rod 3a shows the lowest transmittance degradation in the UV region. 

\subsection{Impact of OH and H$_2$ dopant level on radiation hardness of fused silica}

Different subsets of the BRAN samples were selected to investigate the effects of specific dopants. To simplify comparisons between the chosen subsets, the results at 240 nm, showing the most radiation damage among all the materials, will be used in the following section to discuss the dopants' effect on the radiation hardness of the fused silica. Further comparison at 280 and 320 nm can be found in ~\ref{app:wladdition}.

\begin{figure}[!htbp]
\centering
\includegraphics[width=0.85\textwidth]{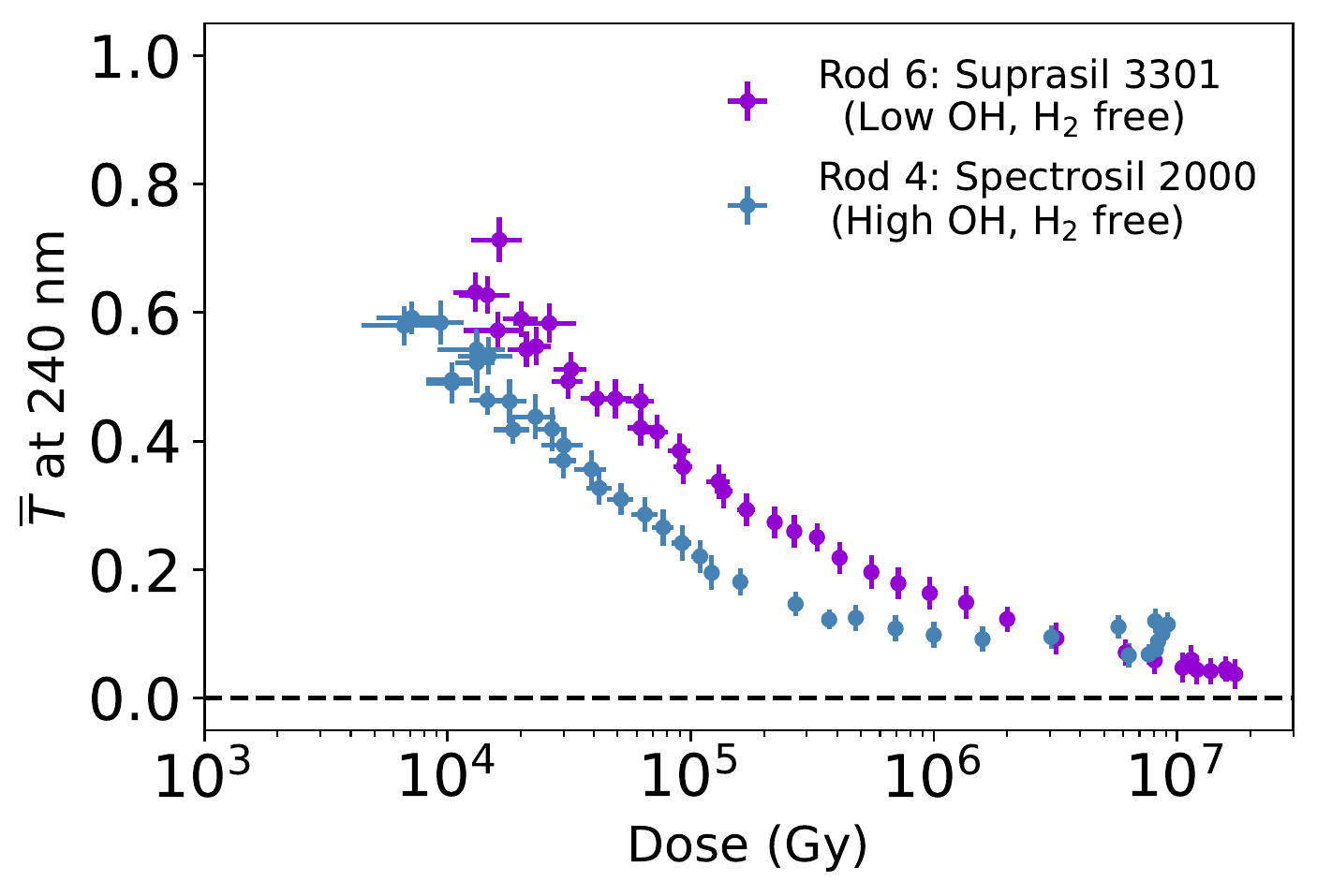}
\caption{Comparison of the transmittance at 240 nm between Rod 6 (Suprasil 3301, Low OH level and H$_2$ free) and Rod 4 (Spectrosil 2000, High OH and H$_2$ free) at different dose levels. A dashed line that corresponds to zero transmittance is drawn for reference.}
\label{fig:fixed_H2_free_changing_OH}
\end{figure}

The OH impact, without the influence of H$_2$, was analyzed by comparing Rod 6 (Suprasil 3301, Low OH) and Rod 4 (Spectrosil 2000, High OH), see Fig.~\ref{fig:fixed_H2_free_changing_OH}. 
A similar transmittance behavior is observed for the two materials. Nevertheless, Suprasil 3301 shows a slower degradation with radiation up to the MGy scale. Both the transmittance of Spectrosil 2000 and Suprasil 3301 seem to reach a plateau in the MGy scale, but exhibits different peculiarities. The first plateaus after 1 MGy of irradiation, with a residual transmittance around 10 \%, while the latter keeps degrading beyond that dose level until reaching a transmittance of a few percent above 10 MGy.

\begin{figure}[!htbp]
\centering
\includegraphics[width=0.85\textwidth]{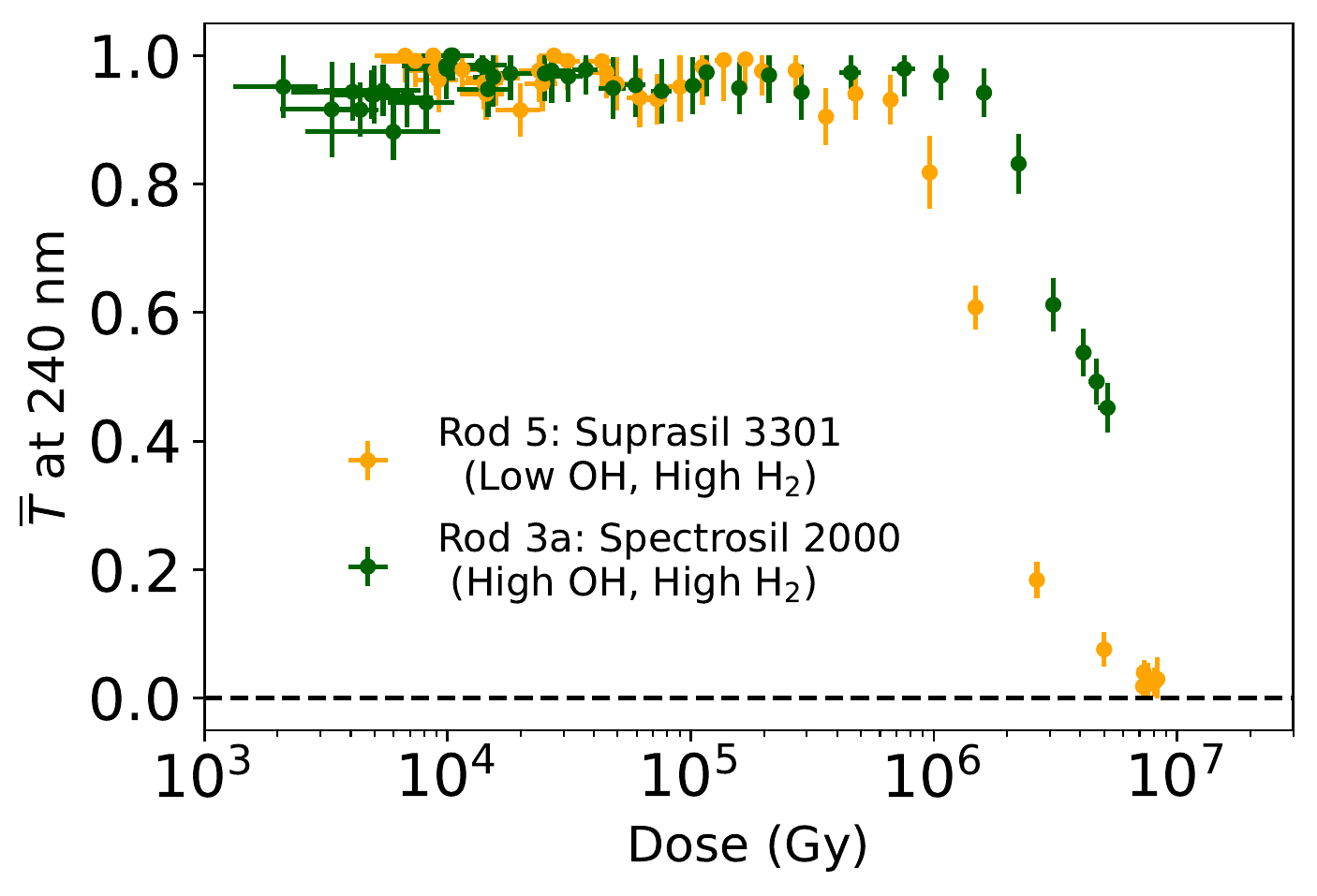}
\caption{Comparison of the transmittance at 240 nm between Rod 5 (Suprasil 3301, Low OH and High H$_2$) and Rod 3a (Spectrosil 2000, High OH and High H$_2$) at different dose levels. A dashed line that corresponds to zero transmittance is drawn for reference.}
\label{fig:fixed_high_H2_changing_OH}
\end{figure}

An analogous study on the impact of OH doping when the fused silica is doped with H$_2$ can be performed by looking at results obtained for Rod 5 
and Rod 3a, 
reported in Fig.~\ref{fig:fixed_high_H2_changing_OH}. Those two rods are characterized by comparable levels of H$_2$, but different concentrations of OH. For doses $>$1 MGy, it is apparent that Rod 3a experienced less transmittance degradation than Rod 5. This trend indicates that a high concentration of OH can help in deferring the degradation when the fused silica is doped with a high concentration of H$_2$.  In general, it appears that the concentration of OH dopant has a much smaller impact on the transmittance degradation compared to the concentration of H$_2$ dopant.

\begin{figure}[!htbp]
\centering
\includegraphics[width=0.85\textwidth]{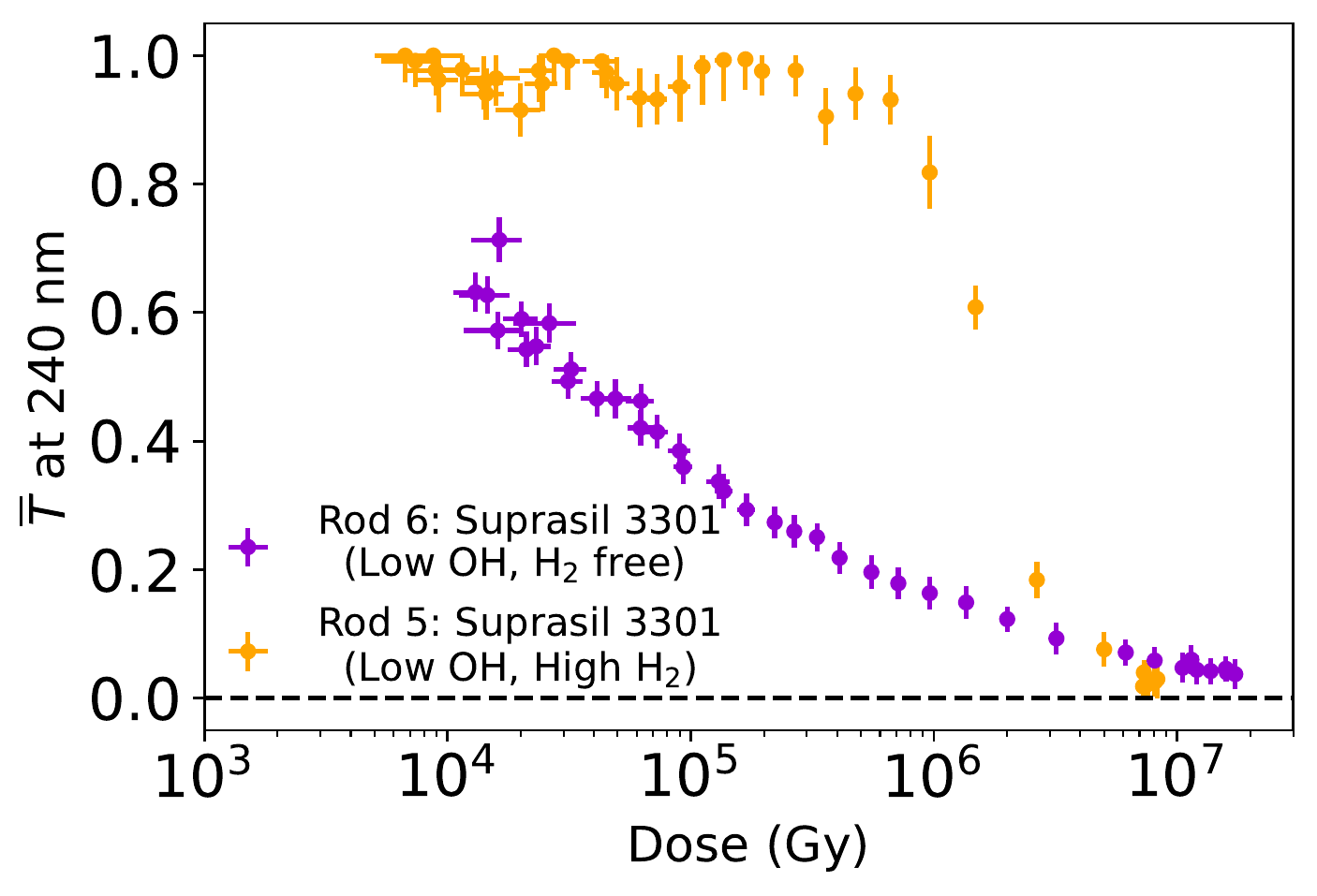}
\caption{Comparison of the transmittance at 240 nm between Rod 6 (Suprasil 3301, Low OH level and H$_2$ free) and Rod 5 (Suprasil 3301, Low OH and High H$_2$) at different dose levels. A dashed line that corresponds to zero transmittance is drawn for reference.}
\label{fig:fixed_low_OH_changing_H2}
\end{figure}

To further support this observation, one can study the impact of H$_2$ by comparing the results for Rod 5 (High H$_2$) and Rod 6 (H$_2$ free) in 
Fig.~\ref{fig:fixed_low_OH_changing_H2}. After 10 MGy of irradiation, the optical transmission of 240 nm light is reduced to a few \% for both materials. However, it is interesting to note that the behavior of the transmittance degradation as a function of received dose varies based on the dopant. Rod 6 exhibits a gradual loss starting from the kGy scale, while Rod 5 shows little transmittance loss up to 500 kGy, but then rapidly degrades and reaches values compatible with a transmittance of zero around 10 MGy, while the H$_2$ free rod tends to maintain a few \% transmittance even beyond 15 MGy of irradiation.

With the available rods, it was also possible to investigate the impact of H$_2$ concentration on the radiation hardness of high OH fused silica (Suprasil 2000) as a function of dose, see Fig.~\ref{fig:fixed_high_OH_changing_H2}. By comparing Rod 3a (high H$_2$) to Rod 3b (mid H$_2$), it appears that a greater H$_2$ concentration defers the threshold for the transmittance degradation to a higher radiation level. In the interval between 0.1 and 1 MGy, the fused silica samples with a high H$_2$ level undergo minimal transparency losses, while the optical transmittance of mid H$_2$ fused silica sample already decays below a few \%. These results suggest that increasing the concentration of H$_2$ by a factor of 4 helps defer the dose turn-on value for the the transmittance degradation by two orders of magnitude.
Conversely, Rod 4, characterized by the absence of H$_2$ doping, exhibits gradual optical losses starting from the kGy range, and its optical transmission remains approximately 0.1 between 0.1 MGy and 10 MGy. It is worth noting that Rod 3b, doped with an intermediate level of H$_2$, does not show the same plateau at the end of the steep degradation starting at $\sim$20 kGy, but reaches full opacity at 240 nm around 1 MGy.


\begin{figure}[!htbp]
\centering
\includegraphics[width=0.85\textwidth]{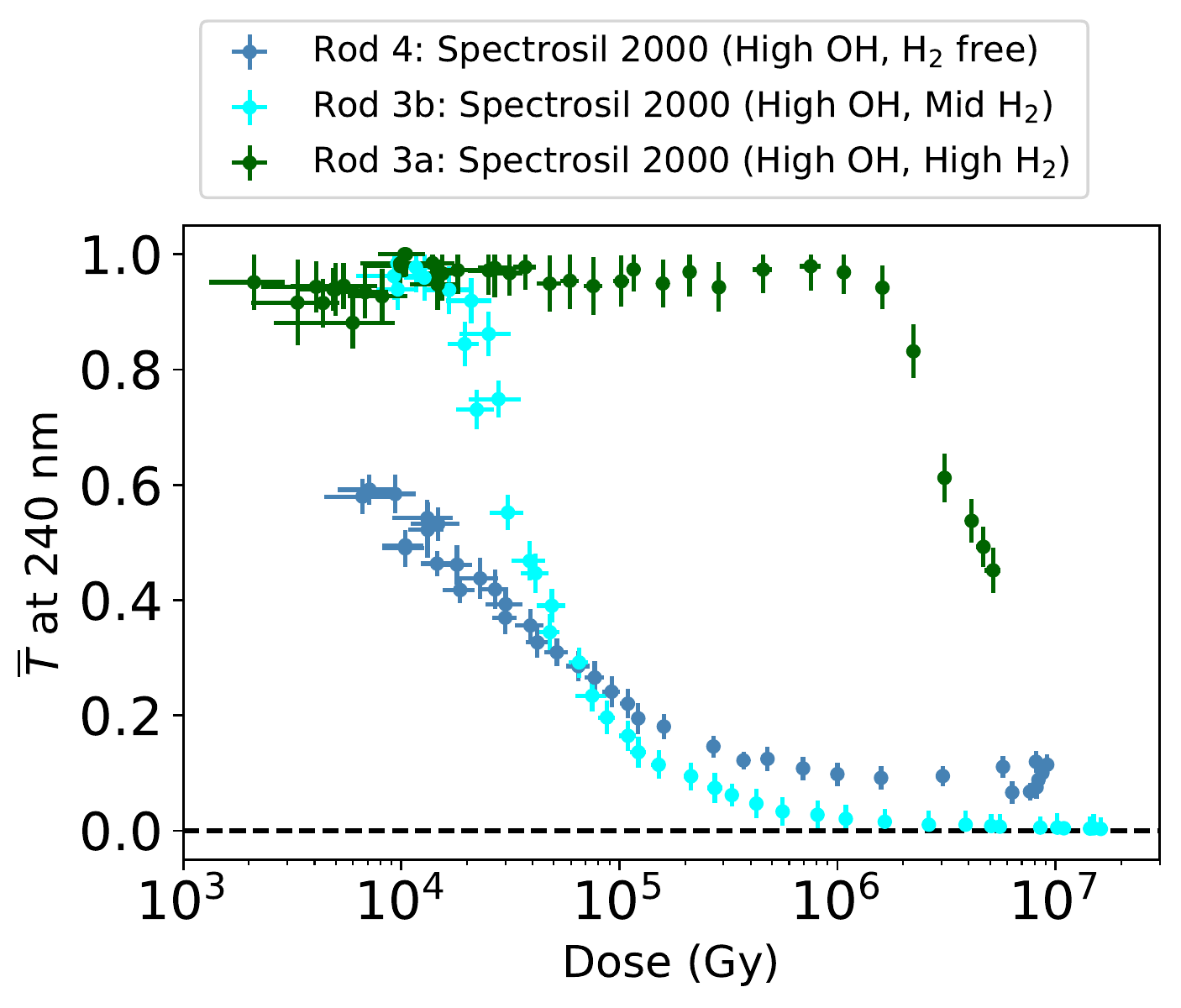}
\caption{Comparison of the transmittance at 240 nm between Rod 4 (Spectrosil 2000, High OH and H$_2$ free), Rod 3b (Spectrosil 2000, High OH and Mid H$_2$), and Rod 3a (Spectrosil 2000, High OH and High H$_2$) at different dose levels.  A dashed line that corresponds to zero transmittance is drawn for reference.}
\label{fig:fixed_high_OH_changing_H2}
\end{figure}

%
%
%
%
\section{Conclusions}
\label{sec:Conclusions}

A BRAN prototype detector containing different types of fused silica rods, doped with various levels of H$_2$ and OH, was irradiated during LHC Run 2. 
The detector was installed in the TAN located in sector 8-1 of the ATLAS long straight section during the 2016-2018 \pp run. Thanks to detailed FLUKA simulations of the accelerator lattice from the ATLAS interaction point up to the TAN, it was possible to evaluate the dose accumulated in the rods. The dose in the rods was found to vary by four orders of magnitude along the vertical direction, allowing for an analysis of the radiation-induced optical attenuation in different fused silica materials across a wide dose range. 

This manuscript presents a full characterization of the optical transmission of irradiated fused silica as a function of wavelength, received dose, and OH and H$_2$ concentration. The irradiation of the fused silica rods was the result of a high energy particles cocktail produced by the showering of very forward neutral particles originating from \pp collisions at IP1. This environment sets our analysis apart from the majority of previous fused silica analyses, where only neutrons, like in a nuclear reactor, or photons, from lasers or other sources, were used to irradiate the materials.

For all the rods analyzed,  most radiation damage appeared in the UV region, with the transmittance losses becoming more severe at lower wavelengths, while very limited transmittance degradation was observed in the wavelength region above 400 nm. 
H$_2$ loading was found to be helpful in countering optical transmission degradation in irradiated fused silica. 
The benefit of H$_2$ doping tends to fade away beyond certain radiation levels depending on H$_2$ and OH concentrations. Once damage in the fused silica starts to appear, the transmittance of H$_2$-doped fused silica degrades faster compared to fused silica without any H$_2$ load, where a saturation trend is observed. Similar observations were made in \cite{bodo:2010} by irradiating different fused silica materials using high energy UV radiation. 

The results presented in this paper highlight the incredible potential of fused silica for optical applications in highly-radioactive environments. The experimental method used for the transmission analysis was developed after irradiation, introducing additional uncertainties due to the sample preparation and the use of the refractive index matching liquid. Additionally, the Run 2 irradiation campaign was limited by the luminosity delivered by the LHC in between 2016 and 2018. 

Run 3 at the LHC offers an invaluable opportunity to carry out a new irradiation campaign thanks to specific grooves obtained in the new BRAN-D detector, that was installed in the TAN at ATLAS and CMS in January 2022. The expected accumulated dose in the samples will surpass the one presented in this paper by at least one order of magnitude, given the higher integrated luminosity that the LHC is planned to deliver in Run 3 and the position of the BRAN-D closer to the shower maximum in the TAN. 
The experience gained during the Run 2 analysis was also exploited to achieve better sample preparation prior to the insertion in the LHC. This effort will streamline the analysis process and help to reduce experimental uncertainties in the transmittance measurement. The extraction of the samples from the LHC is foreseen by the end of Run 3, in 2025.  In the shorter term, an irradiation campaign at the Soreq Nuclear Research Center which is complementary in the total dosage and has a different particle composition will be used for further analysis.

\vspace{0.4cm}
\section*{Acknowledgements}
We would like to thank Dr. Mohammad Amdad Ali, from the Materials Research Laboratory at UIUC, for useful discussions. This work is in part supported by the National Science Foundation, Grants no. PHY-1812377, PHY-1812325 and PHY-2111046, and by the U.S-Israel Binational Science Foundation, Grant no.2020773.

\bibliography{NIMA_bib}

\providecommand{\noopsort}[1]{}\providecommand{\singleletter}[1]{#1}%
\begin{thebibliography}{10}
\expandafter\ifx\csname url\endcsname\relax
  \def\url#1{\texttt{#1}}\fi
\expandafter\ifx\csname urlprefix\endcsname\relax\def\urlprefix{URL }\fi
\expandafter\ifx\csname href\endcsname\relax
  \def\href#1#2{#2} \def\path#1{#1}\fi

\bibitem{milam1998review}
D.~Milam, Review and assessment of measured values of the nonlinear
  refractive-index coefficient of fused silica, Applied optics 37~(3) (1998)
  546--550.

\bibitem{miller2012optical}
S.~Miller, Optical fiber telecommunications, Elsevier, 2012.

\bibitem{marshall1997induced}
C.~D. Marshall, J.~A. Speth, S.~A. Payne, Induced optical absorption in gamma,
  neutron and ultraviolet irradiated fused quartz and silica, Journal of
  non-crystalline solids 212~(1) (1997) 59--73.

\bibitem{skuja2005defects}
L.~Skuja, M.~Hirano, H.~Hosono, K.~Kajihara, Defects in oxide glasses, physica
  status solidi (c) 2~(1) (2005) 15--24.

\bibitem{nurnberg2015bulk}
F.~N{\"u}rnberg, B.~K{\"u}hn, A.~Langner, M.~Altwein, G.~Sch{\"o}tz, R.~Takke,
  S.~Thomas, J.~Vydra, Bulk damage and absorption in fused silica due to
  high-power laser applications, in: Laser-Induced Damage in Optical Materials:
  2015, Vol. 9632, International Society for Optics and Photonics, 2015, p.
  96321R.

\bibitem{brueckner1970properties}
R.~Brueckner, Properties and structure of vitreous silica. {I}, Journal of
  non-crystalline solids 5~(2) (1970) 123--175.

\bibitem{akchurin2003quartz}
N.~Akchurin, R.~Wigmans, Quartz fibers as active elements in detectors for
  particle physics, Review of scientific instruments 74~(6) (2003) 2955--2972.

\bibitem{silica_datasheet}
\href{https://www.heraeus.com/media/media/hca/doc_hca/products_and_solutions_8/optics/Data_and_Properties_Optics_fused_silica_EN.pdf}{Data
  and properties optics fused silica}.
\newline\urlprefix\url{https://www.heraeus.com/media/media/hca/doc_hca/products_and_solutions_8/optics/Data_and_Properties_Optics_fused_silica_EN.pdf}

\bibitem{weeks1960trapped}
R.~Weeks, C.~Nelson, Trapped electrons in irradiated quartz and silica: {II},
  electron spin resonance, Journal of the American Ceramic Society 43~(8)
  (1960) 399--404.

\bibitem{griscom1991optical}
D.~L. Griscom, Optical properties and structure of defects in silica glass,
  Journal of the Ceramic Society of Japan 99~(1154) (1991) 923--942.

\bibitem{griscom1996gamma}
D.~Griscom, $\gamma$ and fission-reactor radiation effects on the visible-range
  transparency of aluminum-jacketed, all-silica optical fibers, Journal of
  applied physics 80~(4) (1996) 2142--2155.

\bibitem{salem2013transparent}
J.~A. Salem, Transparent armor ceramics as spacecraft windows, Journal of the
  American Ceramic Society 96~(1) (2013) 281--289.

\bibitem{Bilki:8824611}
B.~Bilki, Y.~Onel, Design, construction and commissioning of the upgrade
  radiation damage monitoring system of the cms hadron forward calorimeters,
  in: 2018 IEEE Nuclear Science Symposium and Medical Imaging Conference
  Proceedings (NSS/MIC), 2018, pp. 1--4.
\newblock \href {https://doi.org/10.1109/NSSMIC.2018.8824611}
  {\path{doi:10.1109/NSSMIC.2018.8824611}}.

\bibitem{akchurin2020cerium}
N.~Akchurin, N.~Bartosik, J.~Damgov, F.~De~Guio, G.~Dissertori, E.~Kendir,
  S.~Kunori, T.~Mengke, F.~Nessi-Tedaldi, N.~Pastrone, et~al., Cerium-doped
  fused-silica fibers for particle physics detectors, Journal of
  Instrumentation 15~(03) (2020) C03054.

\bibitem{abdullin2008design}
S.~Abdullin, V.~Abramov, B.~Acharya, M.~Adams, N.~Akchurin, U.~Akgun,
  E.~Anderson, G.~Antchev, M.~Arcidy, S.~Ayan, et~al., Design, performance, and
  calibration of {CMS} forward calorimeter wedges, The European Physical
  Journal C 53~(1) (2008) 139--166.

\bibitem{evans2008lhc}
L.~Evans, P.~Bryant, {LHC} machine, Journal of instrumentation 3~(08) (2008)
  S08001.

\bibitem{hl_lhc}
\href{https://hilumilhc.web.cern.ch/}{The {HL}-{LHC} project: High luminosity
  {L}arge {H}adron {C}ollider}.
\newline\urlprefix\url{https://hilumilhc.web.cern.ch/}

\bibitem{apollinari2017high}
G.~Apollinari, O.~Br{\"u}ning, T.~Nakamoto, L.~Rossi, High luminosity large
  hadron collider {HL}-{LHC}, arXiv preprint arXiv:1705.08830 (2017).

\bibitem{PhysRevAccelBeams.25.053001}
P.~Santos~D\'{\i}az, F.~S. Galan, R.~De~Maria, E.~Kepes, P.~Steinberg,
  R.~Longo, S.~Mazzoni, F.~Cerutti, M.~Sabate~Gilarte, A.~Infantino,
  B.~Salvant, M.~Murray, O.~Boettcher, J.~Hansen, V.~Baglin, N.~Zelko,
  E.~Bravin, H.~Mainaud, P.~Bestmann, A.~Herty, P.~Fessia, M.~Gonzalez de~la
  Aleja~Cabana,
  \href{https://link.aps.org/doi/10.1103/PhysRevAccelBeams.25.053001}{Mechanical
  and thermal design of the target neutral beam absorber for the
  high-luminosity lhc upgrade}, Phys. Rev. Accel. Beams 25 (2022) 053001.
\newblock \href {https://doi.org/10.1103/PhysRevAccelBeams.25.053001}
  {\path{doi:10.1103/PhysRevAccelBeams.25.053001}}.
\newline\urlprefix\url{https://link.aps.org/doi/10.1103/PhysRevAccelBeams.25.053001}

\bibitem{aad2008atlas}
G.~Aad, E.~Abat, J.~Abdallah, A.~Abdelalim, A.~Abdesselam, B.~Abi, M.~Abolins,
  H.~Abramowicz, E.~Acerbi, B.~Acharya, et~al., The {ATLAS} experiment at the
  {CERN} {L}arge {H}adron {C}ollider, Journal of instrumentation 3 (2008)
  S08003.

\bibitem{chatrchyan2008cms}
S.~Chatrchyan, G.~Hmayakyan, V.~Khachatryan, C.~Collaboration, et~al., The
  {CMS} experiment at the {CERN} {LHC}, Journal of instrumentation 3~(8) (2008)
  S08004.

\bibitem{Matis:2016raz}
H.~S. Matis, M.~Placidi, A.~Ratti, W.~C. Turner, E.~Bravin, R.~Miyamoto, {The
  BRAN luminosity detectors for the {LHC}}, Nucl. Instrum. Meth. A 848 (2017)
  114--126.
\newblock \href {http://arxiv.org/abs/1612.01238} {\path{arXiv:1612.01238}},
  \href {https://doi.org/10.1016/j.nima.2016.12.019}
  {\path{doi:10.1016/j.nima.2016.12.019}}.

\bibitem{jenni2007zero}
P.~Jenni, M.~Nessi, M.~Nordberg, Zero degree calorimeters for {ATLAS}, Tech.
  rep., CM-P00072881 (2007).

\bibitem{grachov2006status}
O.~A. Grachov, M.~Murray, A.~S. Ayan, P.~Debbins, E.~Norbeck, Y.~Onel,
  D.~d’Enterria, C.~Collaboration, Status of zero degree calorimeter for
  {CMS} experiment, in: AIP Conference Proceedings, Vol. 867, American
  Institute of Physics, 2006, pp. 258--265.

\bibitem{atlas2021radiation}
A radiation-hard zero degree calorimeter for {ATLAS} in the {HL}-{LHC} era,
  Tech. rep., ATLAS Collaboration (2021).

\bibitem{FLUKA:web}
{FLUKA CERN website} \url{https://fluka.cern}.

\bibitem{FLUKA:new}
C.~Ahdida, et~al., { New capabilities of the FLUKA multi-purpose code},
  Frontiers in Physics 9 (2021) 788253.
\newblock \href {https://doi.org/10.3389/fphy.2021.788253}
  {\path{doi:10.3389/fphy.2021.788253}}.

\bibitem{FLUKA:old}
G.~Battistoni, et~al., {Overview of the FLUKA code}, Annals of Nuclear Energy
  82 (2015) 10--18.
\newblock \href {https://doi.org/10.1016/j.anucene.2014.11.007}
  {\path{doi:10.1016/j.anucene.2014.11.007}}.

\bibitem{heraeus_datasheet}
\href{https://www.heraeus.com/media/media/hca/doc_hca/products_and_solutions_8/optics/Data_and_Properties_Optics_fused_silica_EN.pdf}{Quartz
  glass for optics data and properties} (2019).
\newline\urlprefix\url{https://www.heraeus.com/media/media/hca/doc_hca/products_and_solutions_8/optics/Data_and_Properties_Optics_fused_silica_EN.pdf}

\bibitem{alia2017lhc}
R.~G. Al{\'\i}a, M.~Brugger, F.~Cerutti, S.~Danzeca, A.~Ferrari, S.~Gilardoni,
  Y.~Kadi, M.~Kastriotou, A.~Lechner, C.~Martinella, et~al., {LHC} and
  {HL}-{LHC}: Present and future radiation environment in the high-luminosity
  collision points and {RHA} implications, IEEE Transactions on Nuclear Science
  65~(1) (2017) 448--456.

\bibitem{Prelipcean:2777059}
D.~Prelipcean, \href{https://cds.cern.ch/record/2777059}{Comparison between
  measured radiation levels and fluka simulations at charm and in the lhc
  tunnel of p1-5 within the r2e project in run 2. vergleich zwischen gemessenen
  strahlungswerten und fluka-simulationen bei charm undim lhc-tunnel von p1-5
  innerhalb des r2e-projekts in run 2}, presented 29 Jul 2021 (July 2021).
\newline\urlprefix\url{https://cds.cern.ch/record/2777059}

\bibitem{yang202222}
S.~Yang, et~al., {$^{22}$Na activation level measurements of fused silica rods
  in the LHC target absorber for neutrals compared to simulations}, Phys. Rev.
  Accel. Beams 25~(9) (2022) 091001.
\newblock \href {http://arxiv.org/abs/2204.01937} {\path{arXiv:2204.01937}},
  \href {https://doi.org/10.1103/PhysRevAccelBeams.25.091001}
  {\path{doi:10.1103/PhysRevAccelBeams.25.091001}}.

\bibitem{ATLASluminosity}
Luminosity public results {R}un2,
  \url{https://twiki.cern.ch/twiki/bin/view/AtlasPublic/LuminosityPublicResultsRun2},
  accessed: 2020-12-23.

\bibitem{index_liquid_datasheet}
\href{https://cargille.com/wp-content/uploads/2018/10/Refractive-Index-Liquid-Series-AA-n-1.4560-at-589.3-nm-and-25\%C2\%AFC.pdf}{Data
  sheet of refractive index liquid series {AA}} (2018).
\newline\urlprefix\url{https://cargille.com/wp-content/uploads/2018/10/Refractive-Index-Liquid-Series-AA-n-1.4560-at-589.3-nm-and-25\%C2\%AFC.pdf}

\bibitem{Cary5000}
\href{https://www.agilent.com/cs/library/brochures/5990-7786EN_Cary-4000-5000-6000i-UV-Vis-NIR_Brochure.pdf}{Agilent
  {C}ary 4000/5000/6000i series {UV}-{VIS}-{NIR} spectrophotometers}.
\newline\urlprefix\url{https://www.agilent.com/cs/library/brochures/5990-7786EN_Cary-4000-5000-6000i-UV-Vis-NIR_Brochure.pdf}

\bibitem{swinehart1962beer}
D.~F. Swinehart, The {B}eer-{L}ambert law, Journal of chemical education 39~(7)
  (1962) 333.

\bibitem{hope2014diffuse}
A.~H{\"o}pe, Diffuse reflectance and transmittance, in: Experimental Methods in
  the Physical Sciences, Vol.~46, Elsevier, 2014, pp. 179--219.

\bibitem{hamamatsu2017photomultiplier}
K.~Hamamatsu~Photonics, Photomultiplier tubes and assemblies for scintillation
  counting \& high energy physics, Web (accessed date September 14 2020): R8055
  R7250 Datasheet (2017).

\bibitem{bodo:2010}
B.Kühn, E.Arnold, H.-D.Witzke, F.-J.Schilling, M.Stamminger, B.Uebbing,
  Boundary conditions for the saturation and amplitude of induced absorption in
  synthetic fused silica, in: 8th Symposium SiO2, Advanced Dielectrics and
  Related Devices, Varenna, Italy, 2010.

\end{thebibliography}

\appendix
\section{Comparison of transmittance of fused silica characterized by different OH and H$_2$ dopant level at 280 and 320 nm}
\label{app:wladdition}
 
\begin{figure}[!htbp]
\centering
\includegraphics[width=0.49\textwidth]{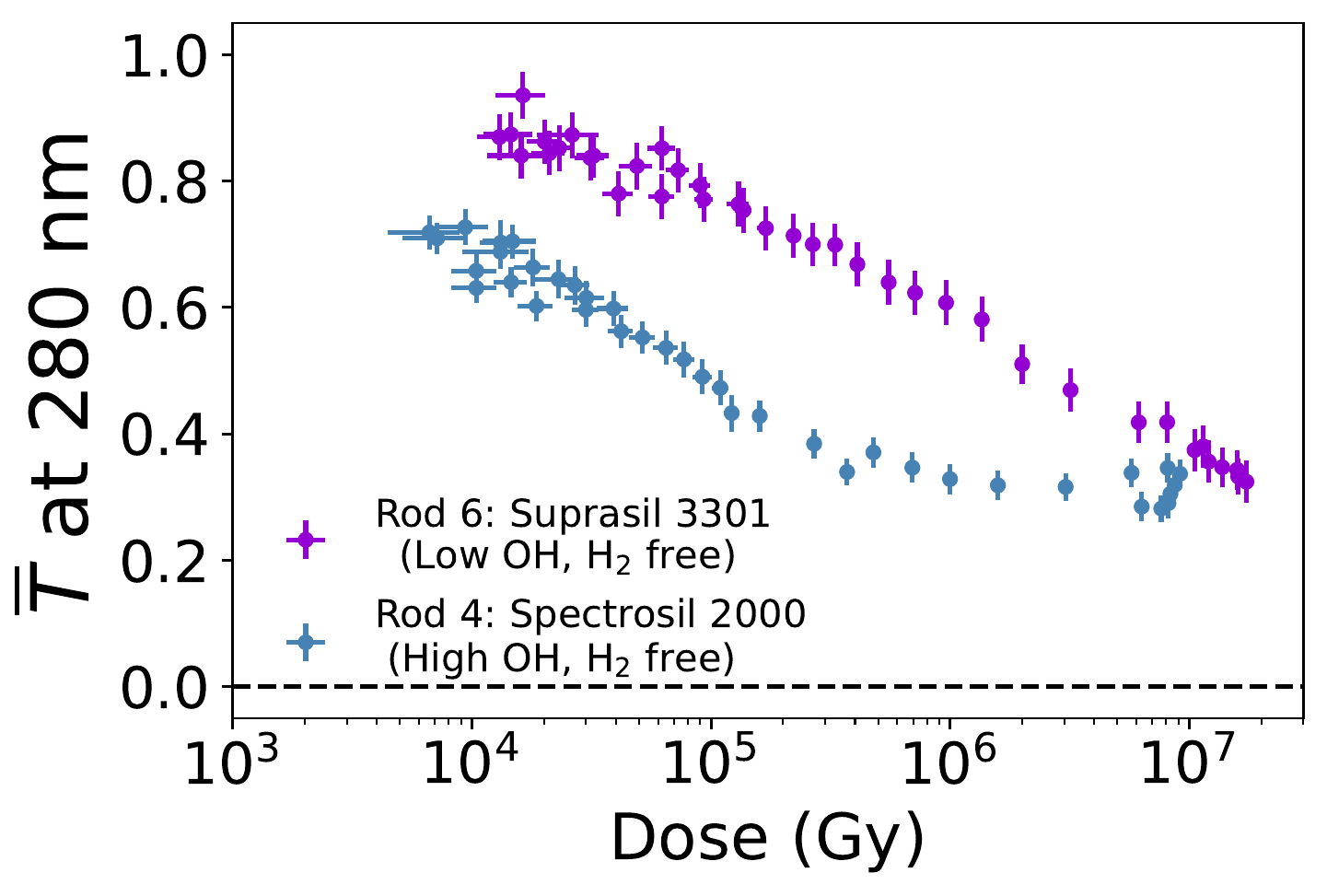}
\includegraphics[width=0.49\textwidth]{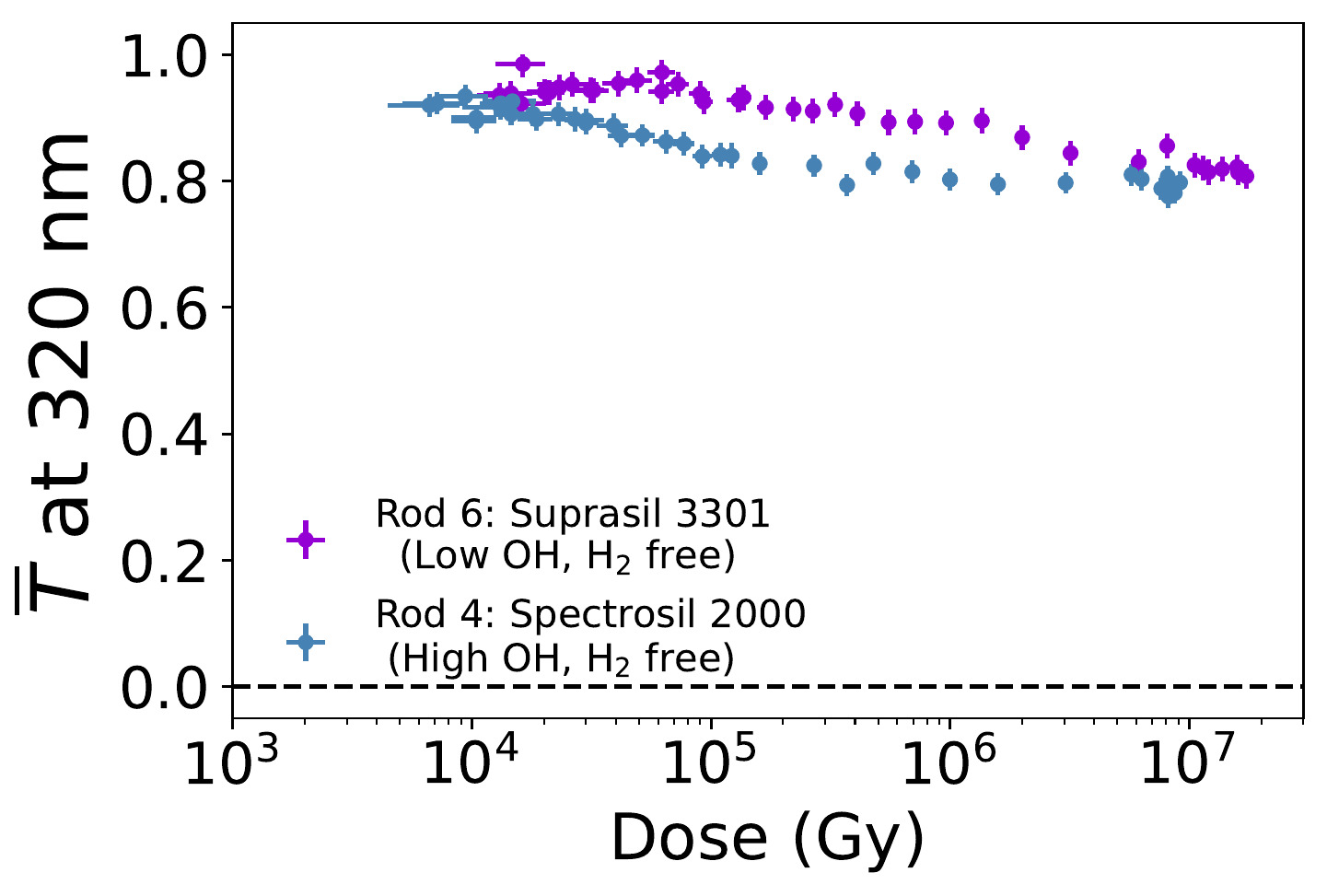}
\caption{Comparison of the transmittance at 280 nm (left) and 320 nm (right) between Rod 6 (Suprasil 3301, Low OH level and H$_2$ free) and Rod 4 (Spectrosil 2000, High OH and H$_2$ free) at different dose levels. A dashed line that corresponds to zero transmittance is drawn for reference.}
\label{fig:fixed_H2_free_changing_OH_280_320}
\end{figure}

\begin{figure}[!htbp]
\centering
\includegraphics[width=0.49\textwidth]{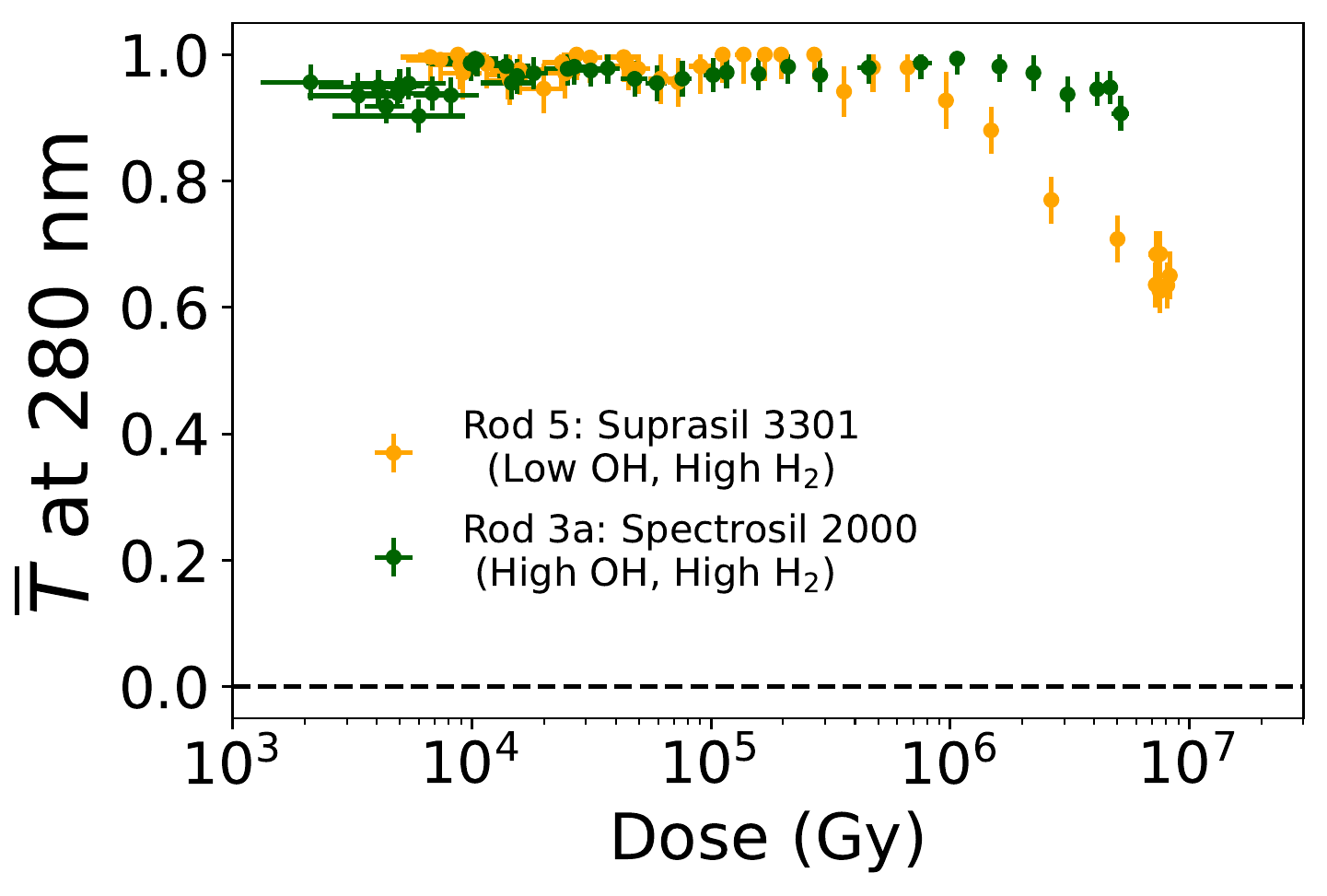}
\includegraphics[width=0.49\textwidth]{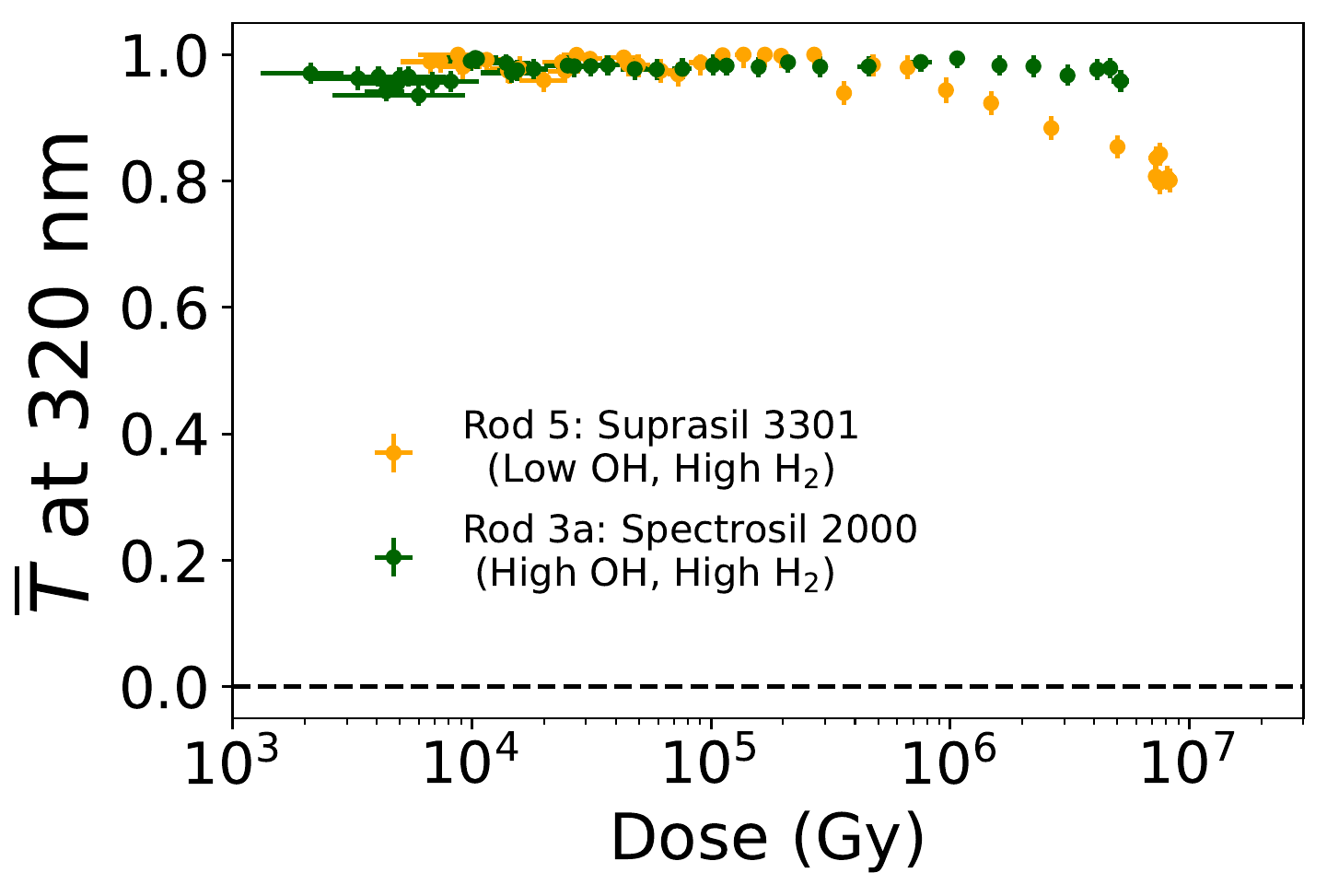}
\caption{Comparison of the transmittance at 280 nm (left) and 320 nm (right) between Rod 5 (Suprasil 3301, Low OH and High H$_2$) and Rod 3a (Spectrosil 2000, High OH and High H$_2$) at different dose levels. A dashed line that corresponds to zero transmittance is drawn for reference.}
\label{fig:fixed_high_H2_changing_OH_280_320}
\end{figure}

\begin{figure}[!htbp]
\centering
\includegraphics[width=0.49\textwidth]{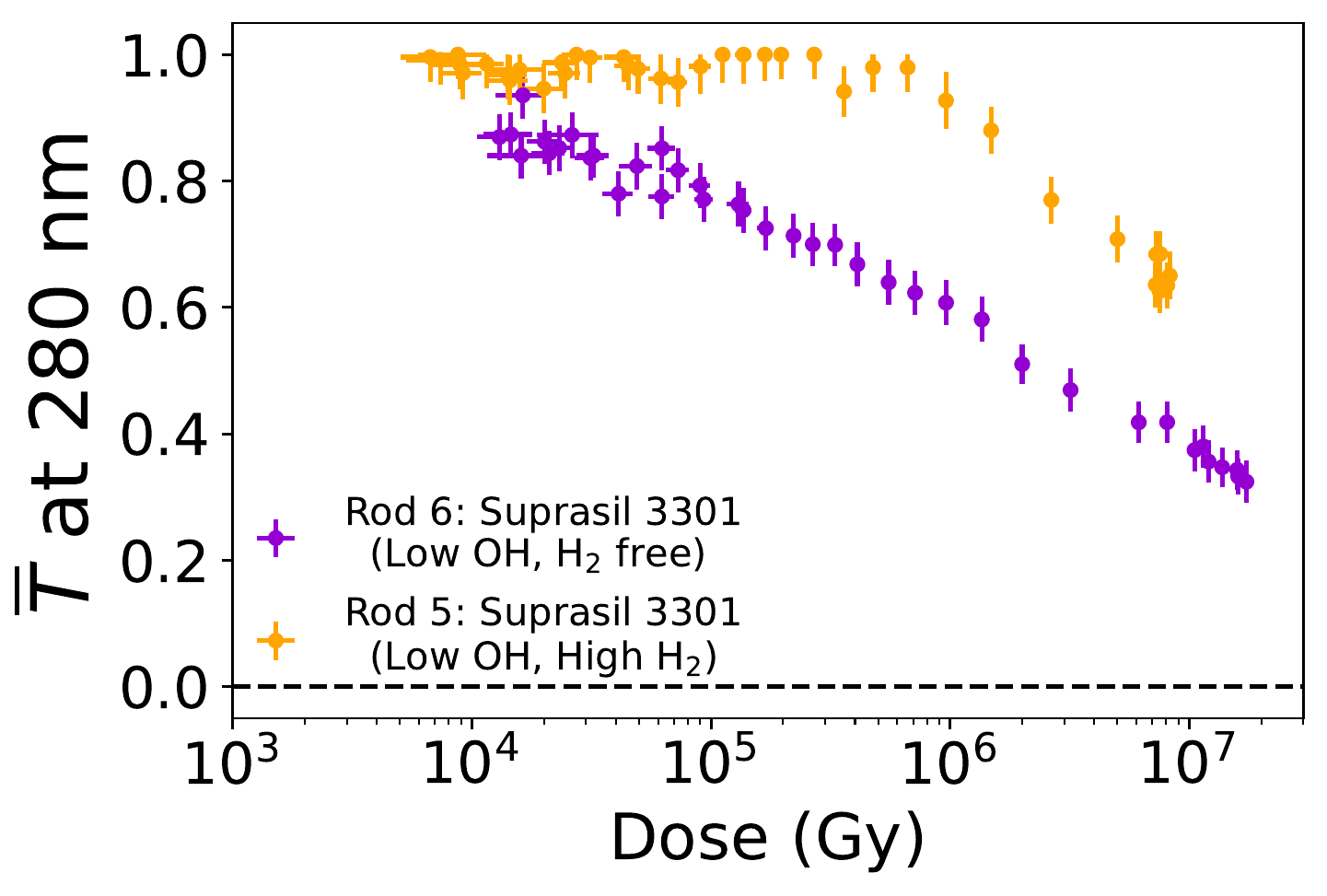}
\includegraphics[width=0.49\textwidth]{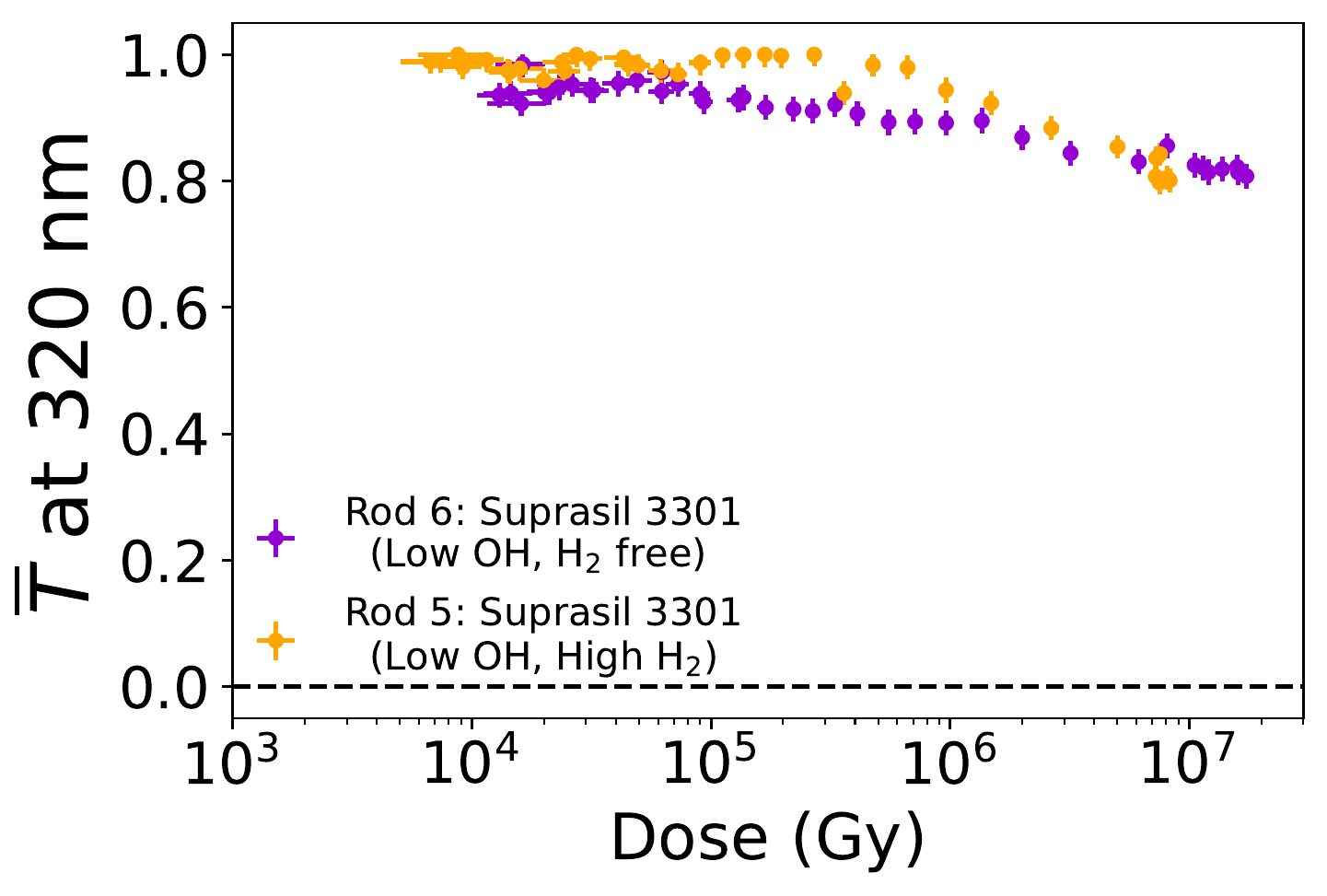}
\caption{Comparison of the transmittance at 280 nm (left) and 320 nm (right) between Rod 6 (Suprasil 3301, Low OH level and H$_2$ free) and Rod 5 (Suprasil 3301, Low OH and High H$_2$) at different dose levels. A dashed line that corresponds to zero transmittance is drawn for reference.}
\label{fig:fixed_low_OH_changing_H2_280_320}
\end{figure}

\begin{figure}[!htbp]
\centering
\includegraphics[width=0.49\textwidth]{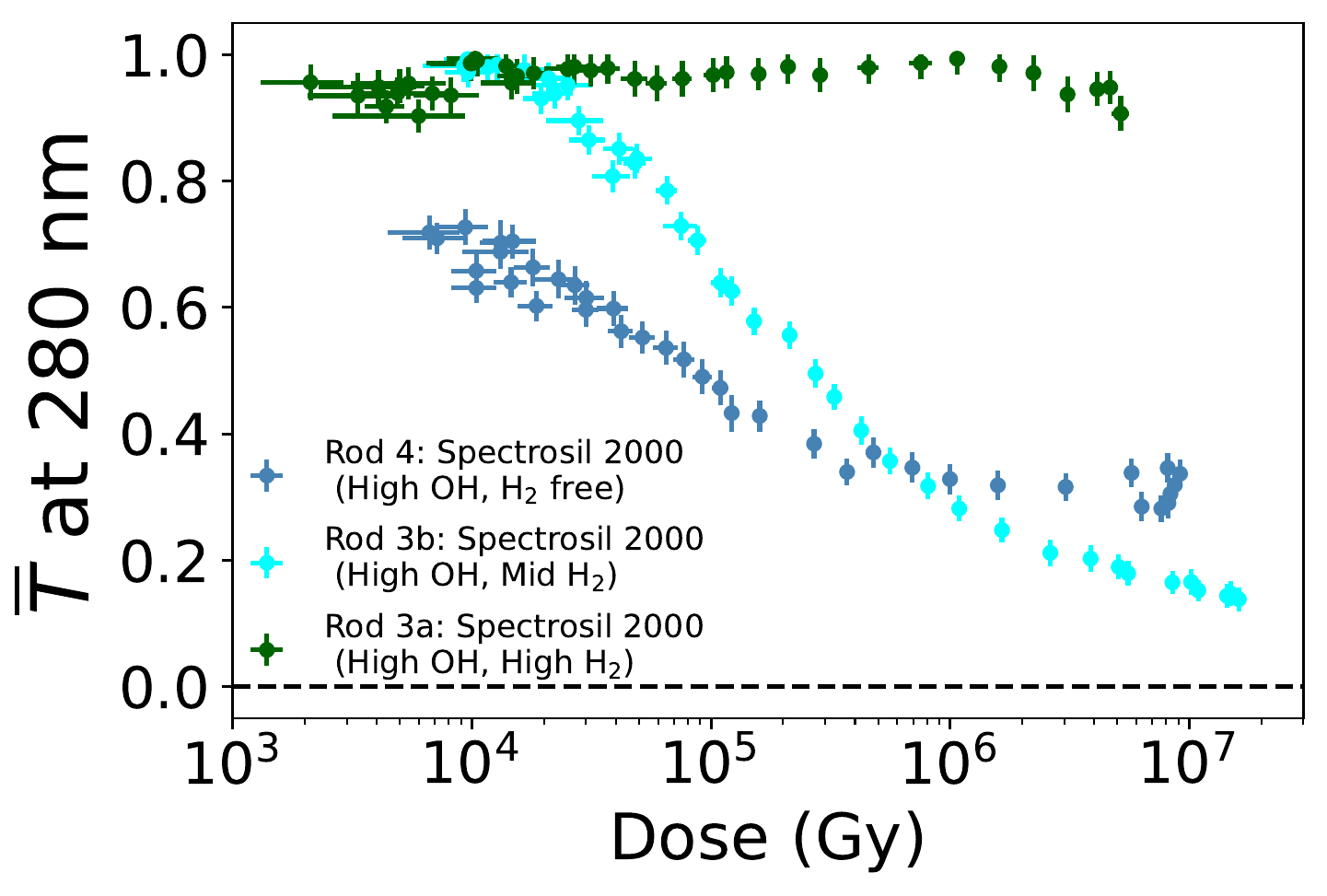}
\includegraphics[width=0.49\textwidth]{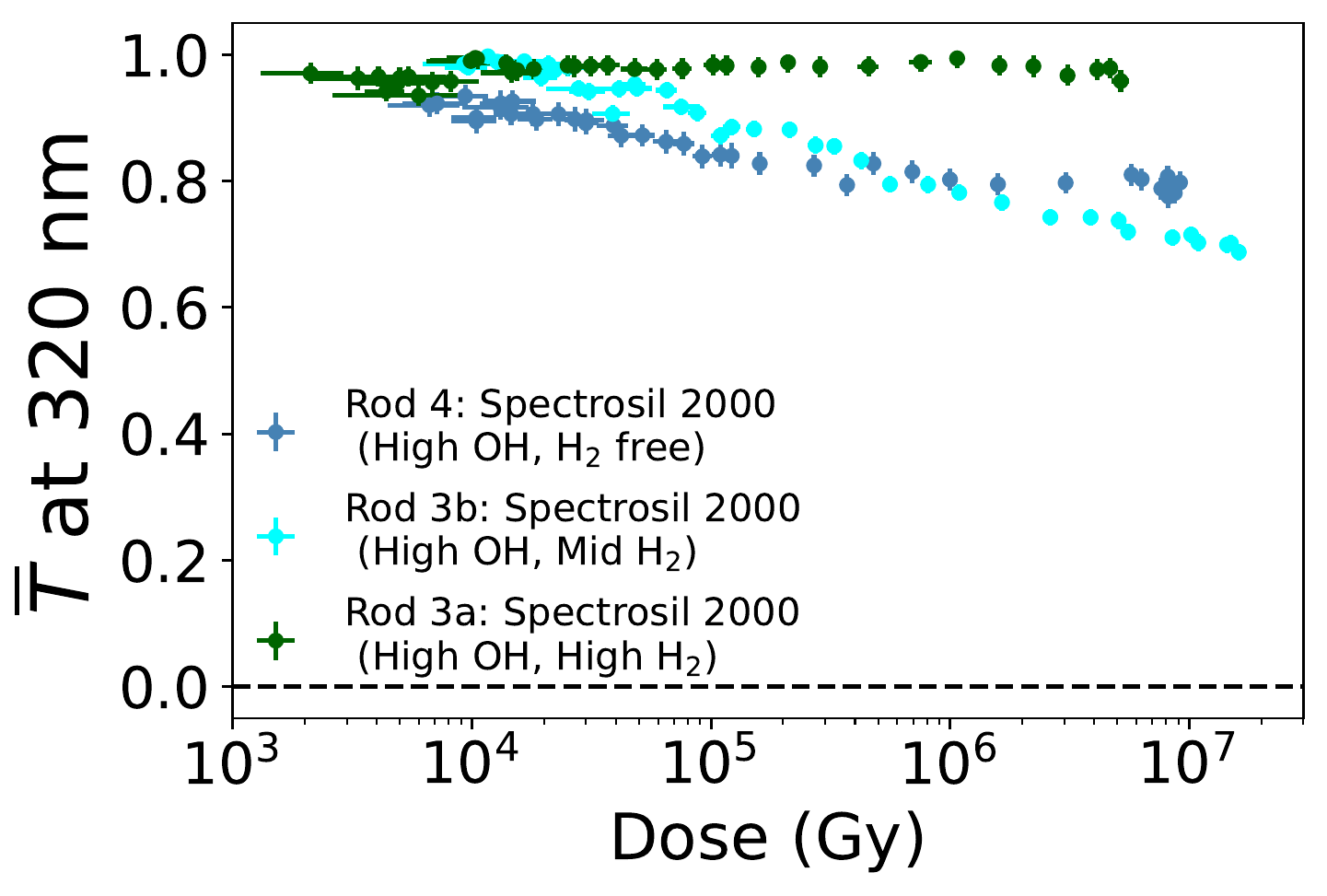}
\caption{Comparison of the transmittance at 280 nm (left) and 320 nm (right) between Rod 4 (Spectrosil 2000, High OH and H$_2$ free), Rod 3b (Spectrosil 2000, High OH and Mid H$_2$), and Rod 3a (Spectrosil 2000, High OH and High H$_2$) at different dose levels.  A dashed line that corresponds to zero transmittance is drawn for reference.}
\label{fig:fixed_high_OH_changing_H2}
\end{figure}

\end{document}